\documentclass[twocolumn]{aastex631}

\usepackage[T1]{fontenc}
\usepackage{dsfont}
\usepackage{xcolor}
\usepackage{comment}
\usepackage{makecell, booktabs}

\newcommand{\wbar}{\bar{w}}

\newcommand{\rp}{r_{\rm p}}
\newcommand{\tomographer}{\textsc{Tomographer}}

\shorttitle{Tomographer}
\shortauthors{Chiang et al.}

\begin{document}

\title{Tomographer: End-to-end Redshift Distribution Estimation for Source Catalogs and Intensity Maps}

\author[0000-0001-6320-261X]{Yi-Kuan Chiang}
\affiliation{Academia Sinica Institute of Astronomy and Astrophysics (ASIAA), No. 1, Sec. 4, Roosevelt Rd., Taipei 10617, Taiwan}

\author[0009-0006-0529-5460]{Yu Voon Ng}
\affiliation{Academia Sinica Institute of Astronomy and Astrophysics (ASIAA), No. 1, Sec. 4, Roosevelt Rd., Taipei 10617, Taiwan}

\author[0009-0009-4749-2149]{Yu-Ren Lin}
\affiliation{Department of Physics, National Taiwan University, No. 1, Sec. 4, Roosevelt Rd., Taipei 10617, Taiwan}
\affiliation{Academia Sinica Institute of Astronomy and Astrophysics (ASIAA), No. 1, Sec. 4, Roosevelt Rd., Taipei 10617, Taiwan}

\author[0000-0002-0636-5698]{Manuchehr Taghizadeh-Popp}
\affiliation{Department of Physics \& Astronomy, Johns Hopkins University, 3400 N. Charles St., Baltimore, MD 21218, USA}
\affiliation{Data Science and AI Institute, Johns Hopkins University, 6225 Smith Ave., Baltimore, MD 21209, USA}

\author[0000-0003-3164-6974]{Brice M\'enard}
\affiliation{Department of Physics \& Astronomy, Johns Hopkins University, 3400 N. Charles St., Baltimore, MD 21218, USA}
\affiliation{Santa Fe Institute, 1399 Hyde Park Rd., Santa Fe, NM 87501, USA}
\affiliation{Anthropic, 500 Howard St., San Francisco, CA 94105, USA}

\begin{abstract}
Redshift information is central to nearly every extragalactic and cosmological application of sky surveys, yet only a small fraction of cataloged sources---and none of the photons forming diffuse backgrounds---have spectroscopic redshifts. Clustering-based redshift inference estimates the redshift distribution of an arbitrary dataset through spatial cross correlation with a reference sample of known redshifts. It relies only on positional information, making it applicable to any tracer of large-scale structure, including source populations and diffuse intensity maps. Its broader adoption, however, has been limited by technical barriers: assembling and characterizing spectroscopic references, expensive pair-counting computations, theoretical corrections, and systematic control. To remove these barriers, we introduce \tomographer, an end-to-end clustering-redshift framework. The key design feature is a set of precomputed ``activation maps'' encoding the spatial pair information of 3~million Sloan Digital Sky Survey spectroscopic galaxies and quasars up to $z\sim4$ over $10{,}000\,{\rm deg}^2$. This eliminates user-end pair counting, reducing computational scaling from $\mathcal{O}(N\log N)$ to $\mathcal{O}(1)$ map multiplications. Given a source catalog or intensity map, \tomographer\ returns the bias-weighted redshift distribution, $b(z)\,{\rm d}N/{\rm d}z(z)$ or $b(z)\,{\rm d}I/{\rm d}z(z)$. We validate the framework against samples with known redshifts, demonstrate accurate uncertainty estimates, and show robustness to survey footprint, spatially varying selection functions, beam smoothing, and foreground contamination. We showcase applications to source catalogs selected by flux, color, photometric redshift, morphology, or variability, and to intensity maps from radio to X-rays. Future \tomographer\ releases will incorporate additional wide-field spectroscopic reference samples as they become available.
\end{abstract}

\section{Introduction} \label{sec:intro}

\begin{figure*}[p]
    \begin{center}
         \includegraphics[width=0.86\textwidth]{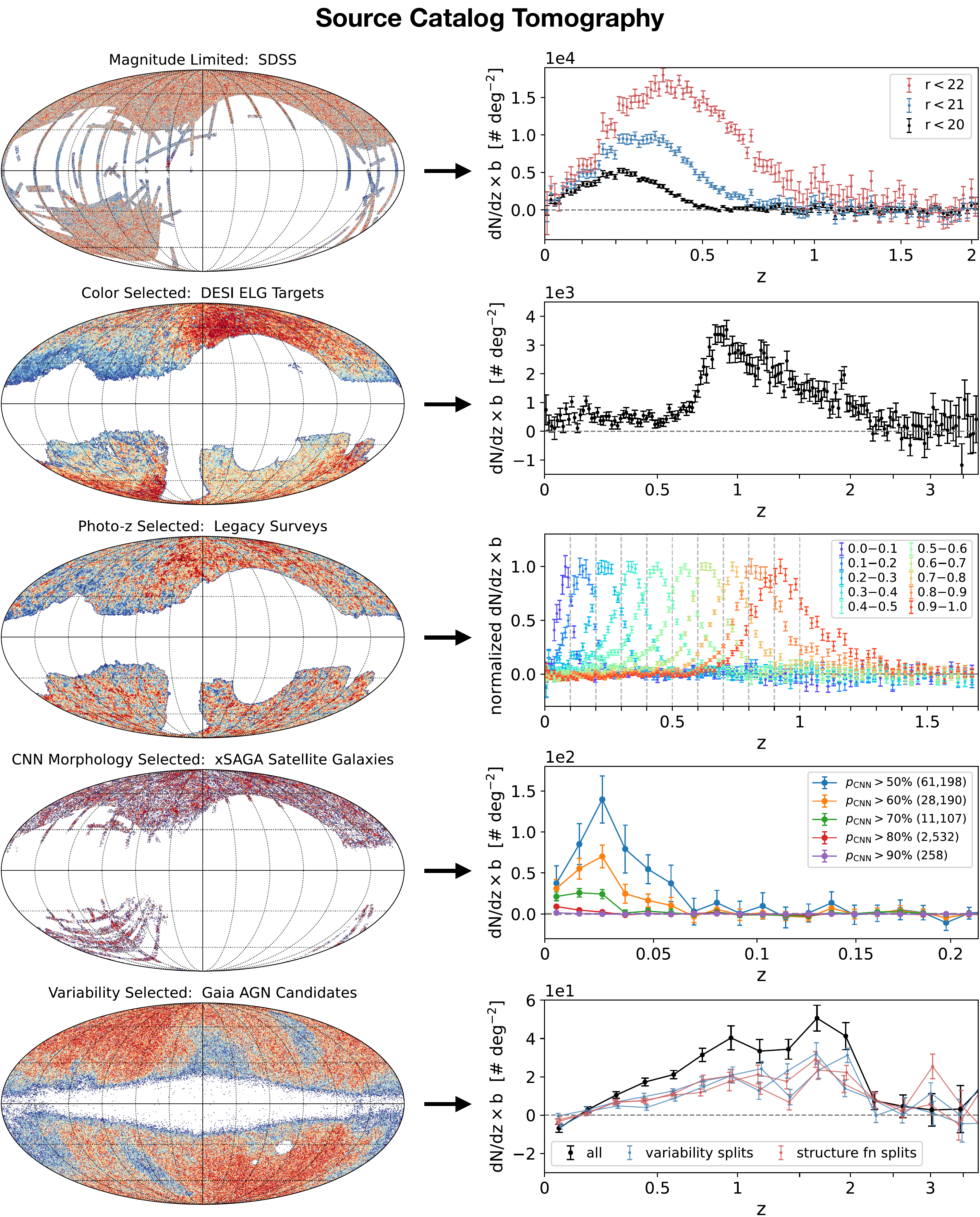}
    \end{center}
    \caption{Source-catalog tomography with \tomographer. Each row shows the input sky density map (left) and recovered bias-weighted redshift distribution (right). From top to bottom: (a) magnitude-limited SDSS photometric galaxies, showing increasing redshift reach with depth; (b) color-selected DESI ELG targets, resolving the $z\sim1$ target population and low-redshift interlopers; (c) Legacy Surveys galaxies in 10 photometric-redshift bins, showing good relative photo-$z$ performance while revealing biases, scatter, and outliers;
    (d) xSAGA low-redshift candidates for satellite-galaxy searches, binned by the CNN-assigned probability of being at $z<0.03$, with higher probabilities recovering progressively lower-redshift populations; and (e) variability-selected Gaia AGN candidates, recovering a broad quasar-like distribution. In every case, only sky positions are used; no photometric or spectral information from the test sample enters the analysis. See Section~\ref{sec:examples} for discussion.}
    
    \label{fig:source_examples}
\end{figure*}

\begin{figure*}[p]
    \begin{center}
         \includegraphics[width=0.86\textwidth]{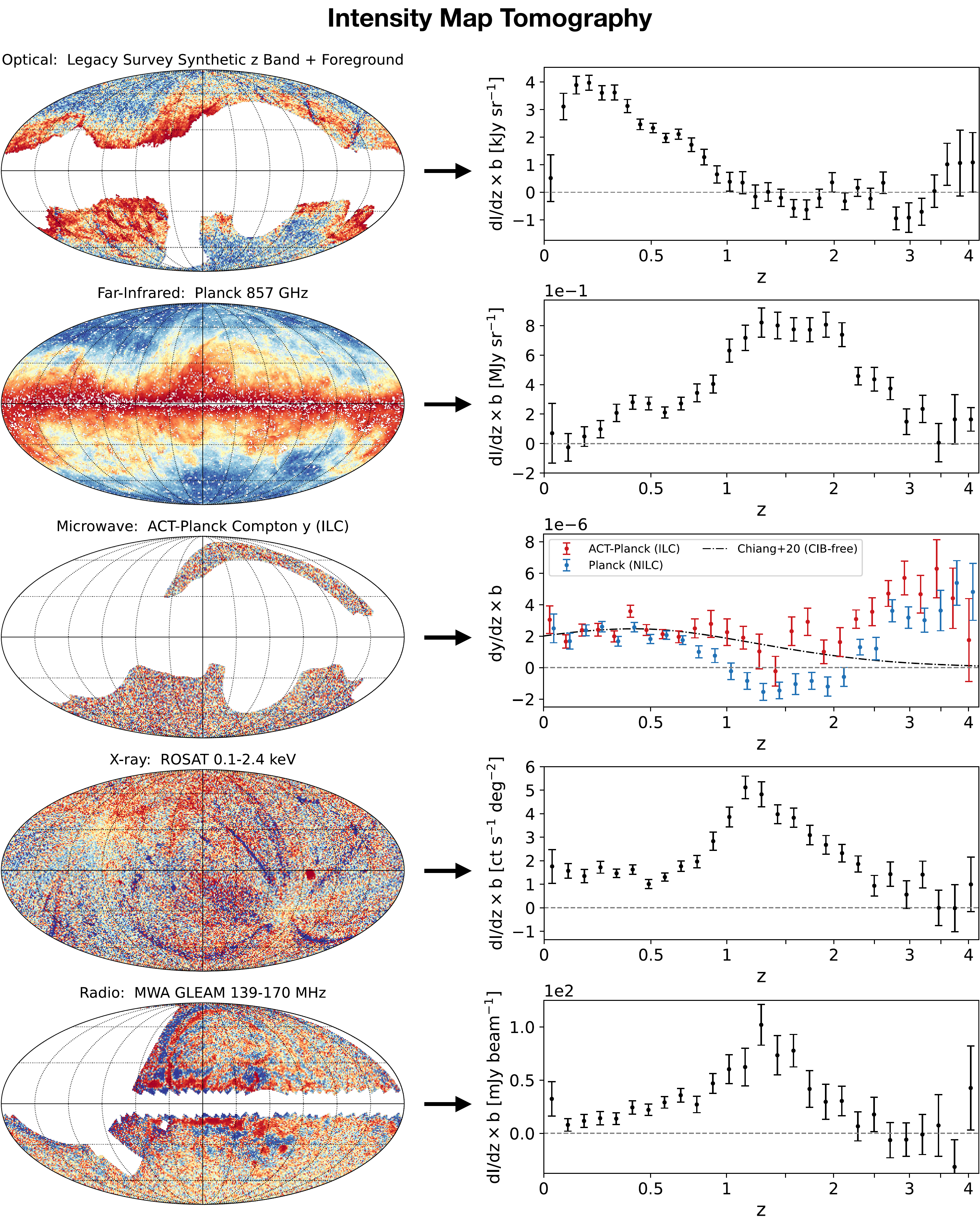}
    \end{center}
    \caption{Intensity-map tomography with \tomographer. Each row shows the input HEALPix intensity map (left) and recovered $b\,({\rm d}I/{\rm d}z)$ (right). From top to bottom: (a) optical: a synthetic Legacy Surveys $z$-band IGL map with added Galactic foregrounds; (b) far-infrared: the Planck 857\,GHz map, recovering the CIB tracing cosmic star formation peaking at $z\sim1$--2; (c) microwave: the ACT-Planck ILC Compton-$y$ map (and Planck NILC), showing the low-redshift thermal SZ effect and high-redshift CIB contamination; (d) X-ray: the ROSAT $0.1$--$2.4$\,keV diffuse map, dominated by AGN at cosmic noon, with a possible low-redshift contribution from hot intracluster and intragroup gas; and (e) radio: the MWA GLEAM 139--170\,MHz map, dominated by radio AGN at $z\sim1$--2, similar to the main X-ray peak. In every case, the redshift decomposition requires no individual source detection and remains robust despite appreciable foregrounds. See Section~\ref{sec:examples} for a discussion.}
    \label{fig:IM_examples}
\end{figure*}

Redshift is the gateway to the third dimension of extragalactic datasets. It converts angular positions into physical scales, fluxes into luminosities, and projected sky maps into a time sequence of cosmic evolution. Modern wide-field surveys detect orders of magnitude more galaxies and other sources than can be followed up spectroscopically. At the same time, much of the information they collect is encoded in unresolved diffuse emission, for which per-source redshifts cannot even be defined. The standard approach, photometric redshift (photo-$z$) estimation, requires multiband photometry of individually detected objects and relies on assumptions about spectral energy distributions (SEDs) or representative machine learning training sets; it degrades for faint, rare, or unusual populations and is inapplicable to intensity maps.

Clustering-based redshift estimation (hereafter clustering redshifts) provides a complementary, purely empirical and statistical route. Since all extragalactic populations trace the same underlying large-scale structure (LSS), the redshift distribution of an arbitrary ``test'' sample can be inferred by measuring its angular cross-correlation with a ``reference'' galaxy sample of known redshifts as a function of reference redshift. The concept was first formalized by \citet{2008ApJ...684...88N}, and later developed by \citet{2010ApJ...721..456M} and \citet{2013MNRAS.433.2857M} using clustering information on linear scales for photo-$z$ \emph{calibration}. \citet{2013arXiv1303.4722M} subsequently generalized the technique for redshift \emph{estimation}, using clustering on smaller, quasi-linear scales to substantially boost the signal-to-noise ratio (SNR) for practical applications \citep{2013MNRAS.431.3307S}. The method has since been validated against spectroscopic samples \citep{2015MNRAS.447.3500R,2016MNRAS.462.1683S} and used to map color to redshift without SED assumptions in photometric surveys such as the Sloan Digital Sky Survey (SDSS) and the Wide-field Infrared Survey Explorer \citep[WISE;][]{2016MNRAS.460..163R,2020JCAP...05..047K,chiang23}. It has also become a key redshift-calibration tool for weak-lensing surveys, including the Dark Energy Survey \citep[DES;][]{2018MNRAS.477.2196D,2018MNRAS.477.1664G,2022MNRAS.510.1223G,2022MNRAS.513.5517C}, the Kilo-Degree Survey \citep[KiDS;][]{2020A&A...642A.200V,hildebrandt21}, the Hyper Suprime-Cam survey \citep{choppin26}, and Euclid \citep{assignies25}, as well as joint analyses \citep{2025arXiv251215963R}. For a broader review of photometric and clustering redshifts, see \citet{2022ARAA..60..363N}.

Because clustering redshift relies only on spatial information, it extends readily from source catalogs to diffuse sky maps, enabling the extragalactic background light (EBL) to be traced across redshift. This extension, known as intensity tomography, has been used to reveal extragalactic imprints in Galactic dust maps \citep{2019ApJ...870..120C}, spectrally tag the evolving UV background \citep{2019ApJ...877..150C}, reconstruct the cosmic infrared background (CIB; \citealp{chiang25}; see also \citealp{schmidt15}), probe the cosmic thermal history through the Sunyaev--Zel'dovich (SZ) effect \citep{2020ApJ...902...56C,2026arXiv260511083S}, characterize the gamma-ray background \citep{2026arXiv260616297K}, and detect the long-predicted cosmic CO and [CII] line backgrounds \citep{chiang26}. In this regime, redshift estimation is performed for \emph{photons} rather than objects: no source detection is required, and the measurement directly probes the radiative history of the Universe rather than the selection function of a galaxy catalog.

While broadly applicable, implementing clustering-redshift estimation remains technically demanding. A typical analysis requires assembling LSS-grade spectroscopic reference catalogs together with the corresponding random catalogs and survey masks, characterizing the reference clustering bias, performing pair counting over millions of objects in fine redshift bins, and guarding against systematics such as spatially varying selection functions and foregrounds before scientific interpretation can begin. Public codes such as \textsc{The-wiZZ}\footnote{\url{https://github.com/morriscb/The-wiZZ}} \citep{2017MNRAS.467.3576M}, \texttt{BallTreeXcorrZ}\footnote{\url{https://github.com/akrolewski/BallTreeXcorrZ}} \citep{2020JCAP...05..047K}, and \texttt{yet\_another\_wizz}\footnote{\url{https://github.com/jlvdb/yet_another_wizz}} \citep{2020A&A...642A.200V} have eased some of these technical demands, but remain expert tools. They also require full spatial pair calculations on the user's end, and none supports intensity maps.

To lower these barriers, we introduce \tomographer, an integrated clustering-redshift framework designed to be fast, robust, and usable end-to-end by non-experts while retaining science-grade accuracy for survey collaborations. \tomographer\ accepts either a source catalog (a list of sky coordinates) or an intensity map in the HEALPix format \citep{2005ApJ...622..759G}, and returns the bias-weighted redshift distribution, $b\,({\rm d}N/{\rm d}z)$ or $b\,({\rm d}I/{\rm d}z)$, in fine redshift bins over $0<z\lesssim4$. A key design feature is the precomputation of all reference-side spatial information into a set of \emph{activation maps}. Once constructed, cross correlations with any test sample reduce to simple inner products between pixel vectors, eliminating user-side pair counting entirely and reducing a full analysis to seconds or minutes. We publicly release this framework as the \tomographer\ software package,\footnote{\url{https://github.com/yuvoonng/tomographer}} accompanying this paper.

\tomographer\ deliberately reports these bias-weighted distributions without assuming the evolution of $b(z)$ needed to recover the underlying redshift distributions. For many LSS applications, $b\,({\rm d}N/{\rm d}z)$ is more fundamental, as it determines cross-correlation signals when the sample traces matter (Section~\ref{subsec:bias_weighted}). Where ${\rm d}N/{\rm d}z$ or ${\rm d}I/{\rm d}z$ itself is needed, additional information or sample subdivision can help reduce or break the bias degeneracy \citep{2013arXiv1303.4722M}.

The breadth of \tomographer\ is showcased in Figures~\ref{fig:source_examples} and \ref{fig:IM_examples}, which offer a preview of ten scientifically diverse applications. The input can consist of sky positions of sources selected by, for example, magnitude, color, photometric redshift, neural network morphology, or variability, or a diffuse intensity map, with examples here spanning wavelengths from radio to X-rays. Across all these datasets, \tomographer\ uses spatial information alone, without relying on photometric or spectral information, to return the bias-weighted redshift distribution in fine redshift bins. We discuss these applications in greater detail in Section~\ref{sec:examples}.

This paper is organized as follows. We review the clustering-redshift formalism in Section~\ref{sec:clustering-z}, then introduce the \tomographer\ algorithm and its precomputed reference-data products in Section~\ref{sec:tomographer}. A series of validation tests is presented in Section~\ref{sec:validation}, and the example applications introduced above are explored in Section~\ref{sec:examples}. A discussion and outlook are presented in Section~\ref{sec:discussion}, followed by a summary in Section~\ref{sec:conclusion}. Throughout, we adopt the \textit{Planck} 2018 ``TT, TE, EE+lowE+lensing'' flat $\Lambda$CDM cosmology \citep{2020A&A...641A...6P}, with $(h,\,\Omega_{\rm c}h^2,\,\Omega_{\rm b}h^2,\,A_{\rm s},\,n_{\rm s}) = (0.6737,\,0.1198,\,0.02233,\,2.097\times10^{-9},\,0.9652)$.

\section{Formalism}
\label{sec:clustering-z}

We begin by introducing a generalized clustering-redshift formalism for recovering the redshift distribution of projected extragalactic fields tracing the LSS. Building on \citet{2013arXiv1303.4722M} and its extension to intensity fields by \citet{2019ApJ...877..150C,chiang25}, we treat source catalogs and intensity maps on equal footing.

\subsection{Projected Observables}
\label{subsec:projected}

We define a generalized projected observable, $S(\phi)$, where $\phi$ denotes the angular coordinate on the sky. Depending on the application, $S(\phi)$ may represent either the continuous surface-density field $N(\phi)$ constructed from a galaxy or source catalog (in units of, e.g., $\#\,\mathrm{deg}^{-2}$), or an intensity map $I(\phi)$ of the EBL retaining its native map units (e.g., MJy\,sr$^{-1}$). More generally, $S(\phi)$ may represent any projected extragalactic field. Source-density and intensity fields primarily differ in how the underlying signal is weighted, e.g., by object counts, photon counts, or energy, and the distinction may be blurred for sparse data such as high-energy photon or particle events. As a projected field, $S$ is the integral of signals along the line of sight:
\begin{equation}
    S(\phi)
    =
    \int
    \frac{\mathrm{d}S(\phi,z)}{\mathrm{d}z}
    \,\mathrm{d}z,
    \label{eq:los_projection}
\end{equation}
where $\mathrm{d}S(\phi,z)/\mathrm{d}z$ denotes the unknown differential 3D contribution from redshift $z$.

A meaningful redshift inference requires $S$ to have a unique, well-defined redshift distribution across the analysis footprint. For sources, this means a consistently selected underlying population; samples selected under different criteria can be combined only if they share the same footprint. Perhaps counterintuitively, intensity maps generally satisfy this homogeneity requirement naturally after photometric calibration, despite often being contaminated by foregrounds. Such contamination is a systematic to mitigate rather than a violation of the underlying assumption. Unlike source catalogs, the EBL has no imposed selection function beyond the observing wavelength and optional masking. Its redshift distribution is therefore directly physically interpretable.

For clustering analyses, we define the excess field
\begin{equation}
    \Delta S(\phi)
    \equiv
    S(\phi)-\langle S\rangle,
    \label{eq:deltaS}
\end{equation}
where $\langle S\rangle$ denotes the ensemble average of $S$, or equivalently its wide-area mean. We choose not to normalize $\Delta S$ into a fractional overdensity. It therefore retains the physical units of $S$ and is less sensitive to additive contamination, such as foreground stars in source catalogs or Milky Way emission in intensity maps, both of which can bias conventional density-contrast normalization.

To probe the redshift distribution of $S$, the clustering-based method requires a clean, spectroscopically confirmed reference sample tracing the underlying LSS in 3D with known redshifts. For each narrow redshift slice $z_i$, we define the projected overdensity field 
\begin{equation}
    \delta_r(\phi,z_i)
    =
    \frac{N_r(\phi,z_i)-\langle N_r(z_i)\rangle}
         {\langle N_r(z_i)\rangle},
    \label{eq:deltar}
\end{equation}
 where $N_r$ and $\langle N_r\rangle$ are the local and mean reference surface number densities, respectively. Unlike  $\Delta S$, $\delta_r$ is normalized and dimensionless. Throughout, we use the subscripts $t$ and $r$ to denote quantities associated with the test and reference fields, respectively; $S(\phi)$ always refers to the test field with the subscript $t$ omitted for brevity.

\subsection{Redshift Deprojection}
\label{subsec:deprojection}

Clustering-redshift estimation aims to recover the mean redshift distribution of the test observable $S$,
\begin{equation}
    \frac{\mathrm{d}S}{\mathrm{d}z}(z)
    \equiv
    \left\langle
    \frac{\mathrm{d}S(\phi,z)}{\mathrm{d}z}
    \right\rangle,
    \qquad
    \langle S \rangle
    =
    \int
    \frac{\mathrm{d}S}{\mathrm{d}z}(z)
    \,\mathrm{d}z,
    \label{eq:dSdz}
\end{equation}
where $\langle\cdot\rangle$ denotes an ensemble or footprint average, and the second relation projects over redshift to recover the normalization $\langle S\rangle$. Since $z$ is dimensionless, $\mathrm{d}S/\mathrm{d}z$ carries the same units as $S$.

Although sky projection largely erases redshift information, $\mathrm{d}S/\mathrm{d}z$ can be estimated statistically by the scale-averaged angular cross correlation,
\begin{equation}
    \bar{w}_{tr}(z_i)
    \equiv
    \langle
    \Delta S \cdot \delta_r(z_i)
    \rangle,
    \label{eq:w_tr_definition}
\end{equation}
between the 2D test field and the 3D references in each radial bin $z_i$. Since $\delta_r$ is dimensionless while $\Delta S$ is left unnormalized, $\bar{w}_{tr}$ retains the physical units of $S$.

Under the assumption that tracer fluctuations are linearly related to the matter density field, the tomographically measured clustering amplitudes satisfy
\begin{equation}
    \bar{w}_{tr}(z_i)
    =
    \bar{w}_{\rm m}(z_i)\,
    b_t(z_i)\,
    b_r(z_i)\,
    \frac{\mathrm{d}S}{\mathrm{d}z}(z_i),
    \label{eq:master}
\end{equation}
where $b_t$ and $b_r$ denote the clustering bias factors of the test field and reference sample, respectively, and $\bar{w}_{\rm m}$ is the scale-averaged angular cross correlation between the matter density field in the thin reference slice and its projected counterpart. The bias factors generally vary with redshift and may also depend mildly on scale. We neglect the typically weak effects of gravitational lensing. Under the Limber approximation \citep{1953ApJ...117..134L}, the angular matter correlation function is given by
\begin{equation}
    w_{\rm m}(\theta,z_i)
    =
    \frac{H(z_i)}{c}
    \int_0^\infty
    \frac{k\,\mathrm{d}k}{2\pi}\,
    P_{\rm m}^{\rm NL}(k,z_i)\,
    J_0\!\left[k\chi(z_i)\theta\right],
    \label{eq:wm}
\end{equation}
where $P_{\rm m}^{\rm NL}$ is the nonlinear matter power spectrum for our assumed cosmology, $\chi$ is the comoving distance, and $J_0$ is the zeroth-order Bessel function.

In $\bar{w}_{tr}$ and $\bar{w}_{\rm m}$, the overbar denotes a scale integral that combines the corresponding angular correlation function into an effective clustering amplitude,
\begin{equation}
    \bar{w}
    =
    \int_{\theta_{\rm min}}^{\theta_{\rm max}}
    W(\theta)\,
    w(\theta)\,
    \mathrm{d}\theta,
    \label{eq:weighting}
\end{equation}
where $(\theta_{\rm min},\theta_{\rm max})$ define the chosen clustering scale, together with a power-law weighting function, $W(\theta)=\theta^\gamma$. We adopt $\gamma=-0.8$, which approximately matches the typical shape of $w(\theta)$ and thus substantially enhances the SNR by upweighting small-scale clustering, following the same principle as matched filtering \citep[e.g.,][]{connolly02, wang13}.

The reference bias $b_r$ can be viewed as a correction term in Equation~(\ref{eq:master}) that accounts for the clustering properties of the chosen reference sample. After correcting for $b_r$, different reference tracers (e.g., galaxies or quasars) should yield the same tomographic redshift measurement of the test field, a behavior that follows directly from the linearity of the equation itself and is empirically validated in Section~\ref{subsec:val_bref}. We measure $b_r$ from the dimensionless reference autocorrelation $\bar{w}_{rr}\equiv\langle\delta_r\cdot\delta_r\rangle$, which for redshift bins wider than the radial correlation length gives
\begin{equation}
    b_r
    \approx
    \sqrt{\frac{\bar{w}_{rr}}{\bar{w}_{\rm m}'}},
    \qquad
    \bar{w}_{\rm m}'
    =
    \frac{\bar{w}_{\rm m}}{\Delta z},
    \label{eq:b_r_simple}
\end{equation}
where $\bar{w}_{\rm m}'$ accounts for the dilution of the binning-independent $\bar{w}_{\rm m}$ in Equation~(\ref{eq:wm}) over a finite redshift bin $\Delta z$. For narrower redshift bins approaching the radial correlation length, however, correlated reference pairs can leak across adjacent slices. We account for this empirically by measuring the angular reference correlations $\bar{w}_{rr}^{ij}$ within each bin and with its two neighboring bins. The resulting ``tri-band'' corrected reference bias is
\begin{equation}
    b_r(z_i)
    =
    \left[
    \sum_{j=i-1}^{i+1}
    \frac{
        \Delta z_j\,\bar{w}_{rr}^{ij}
    }{
        \bar{w}_{\rm m}(z_j)
    }
    \right]^{1/2}.
    \label{eq:b_r}
\end{equation}

Together, Equations~(\ref{eq:master})--(\ref{eq:b_r}) define the basic formalism. In practice, correcting the measured $\bar{w}_{tr}$ for the empirically determined reference bias $b_r$ and the theoretically computed matter-clustering amplitude $\bar{w}_{\rm m}$ yields the product $b_t(z)\,\mathrm{d}S/\mathrm{d}z(z)$ as the directly observable quantity. The redshift distribution $\mathrm{d}S/\mathrm{d}z$ therefore remains degenerate with the generally unknown test-field bias evolution $b_t(z)$. Hereafter, we write $b(z)$ for $b_t(z)$ unless otherwise specified. We discuss this bias degeneracy in Section~\ref{sec:discussion}.

\subsection{Correlation Estimators}
\label{subsec:estimator}

With data, the clustering amplitudes $\bar{w}_{tr}$ and $\bar{w}_{rr}$ are measured using pair-counting estimators. A continuous field such as $\Delta S(\phi)$ is represented by discrete pixels, with each pixel treated analogously to an object and weighted by its field value. If the fields were expressed as normalized density contrasts, the standard choices would be the Landy--Szalay and Davis--Peebles estimators \citep{1983ApJ...267..465D, 1993ApJ...412...64L}. In our formulation, however, the unnormalized $\Delta S(\phi)$ leads to an unnormalized two-point function (Equation~(\ref{eq:w_tr_definition})), motivating modified estimators in \tomographer\ that preserve the physical units of $S$ in $\bar{w}_{tr}$.

When a test random catalog is available, we adopt an unnormalized Landy--Szalay-like estimator,
\begin{equation}
    \bar{w}_{tr}
    =
    D_tD_r
    -
    D_tR_r
    -
    R_tD_r
    +
    R_tR_r,
    \label{eq:w_tr_ls}
\end{equation}
whereas otherwise we use an unnormalized Davis--Peebles-like estimator,
\begin{equation}
    \bar{w}_{tr}
    =
    D_tD_r
    -
    D_tR_r.
    \label{eq:w_tr_dp}
\end{equation}
Here, $D$ and $R$ denote data and random samples, with randoms normalized to the corresponding data sample by their relative catalog sizes; $t$ and $r$ refer to the test and reference fields as before. Thus, for example, $D_tD_r$ denotes the test-data--reference-data pair count, generalized here as a weighted pair accumulation. These accumulations include field-value weights (e.g., pixelized $\Delta S(\phi)$ or $N_r(\phi)$), the angular weighting of Equation~(\ref{eq:weighting}), and any additional systematics or SNR weights, yielding an integrated, one-bin clustering measurement. For unweighted source catalogs, they reduce to ordinary pair counts.

Unlike $\bar{w}_{tr}$, the reference autocorrelation $\bar{w}_{rr}$ is dimensionless and is measured using the standard Davis--Peebles form, applied to our scale-integrated, one-bin clustering measurement,
\begin{equation}
    \bar{w}_{rr}
    =
    \frac{D_rD_r}
         {D_rR_r}
    -1,
    \label{eq:w_rr_dp}
\end{equation}
The resulting $\bar{w}_{rr}$ is used to determine $b_r$ through Equation~(\ref{eq:b_r}).

Having defined the clustering estimators underlying the redshift inference of test sample $b\,\mathrm{d}S/\mathrm{d}z$, we turn in Section~\ref{sec:tomographer} to their implementation in \tomographer, including an efficient algorithm for evaluating these estimators in Section~\ref{subsec:algorithm}.

\begin{figure*}[t!]
    \begin{center}
         \includegraphics[width=0.88\textwidth]{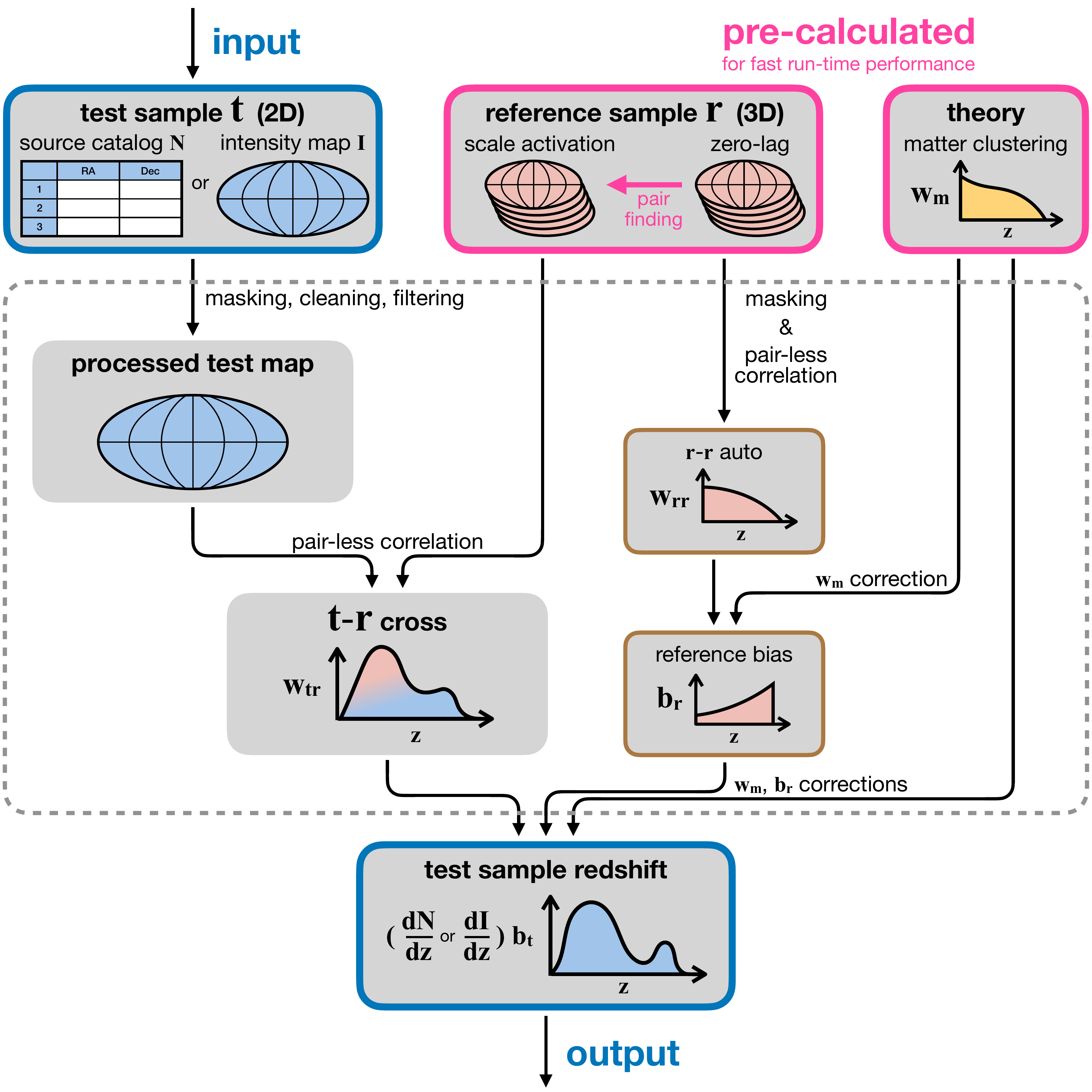}
    \end{center}
\caption{Schematic of the \tomographer\ pipeline. The upper blue boxes denote user inputs: the test sample, either a source catalog or intensity map, and its footprint or random catalog (Section~\ref{subsec:config}). The lower blue box shows the desired bias-weighted redshift distribution, $b\,({\rm d}N/{\rm d}z)$ or $b\,({\rm d}I/{\rm d}z)$ (Section~\ref{subsec:deprojection}). To recover these distributions, the test--reference clustering $\wbar_{tr}(z)$ provides the main redshift deprojection signal, $\wbar_{rr}(z)$ determines the reference bias $b_r(z)$, and $\wbar_{\rm m}(z)$ anchors the matter-clustering amplitude, as summarized by Equation~(\ref{eq:master}). \tomographer\ minimizes user-facing technical barriers and runtime by fixing the reference sample (Section~\ref{subsec:reference}) and precomputing its spatial information. Pink boxes denote precomputed products, including $\wbar_{\rm m}(z)$ (Section~\ref{subsec:scales}) and per-redshift activation maps (Section~\ref{subsec:algorithm}), enabling fast, ``pair-less'' correlation measurements. Tan boxes denote the runtime measurement of $\wbar_{rr}(z)$ and the resulting reference bias $b_r(z)$ over the overlapping footprint.}
    \label{fig:flow_chart}
\end{figure*}

\section{Tomographer}\label{sec:tomographer}

We designed \tomographer\ as an integrated software and data framework that minimizes user-facing processing through reusable precomputed data products and built-in implementations of technical analyses. This design is possible because wide-field clustering-redshift analyses generally rely on a fixed, well-characterized spectroscopic reference sample. Users need only supply the test sample as the general observable $S$ together with its random catalog (or survey mask) and select any optional analysis settings, after which \tomographer\ automatically performs the end-to-end processing. The overall framework is illustrated in Figure~\ref{fig:flow_chart}.

\begin{figure*}
    \centering
    \includegraphics[width=0.995\textwidth]{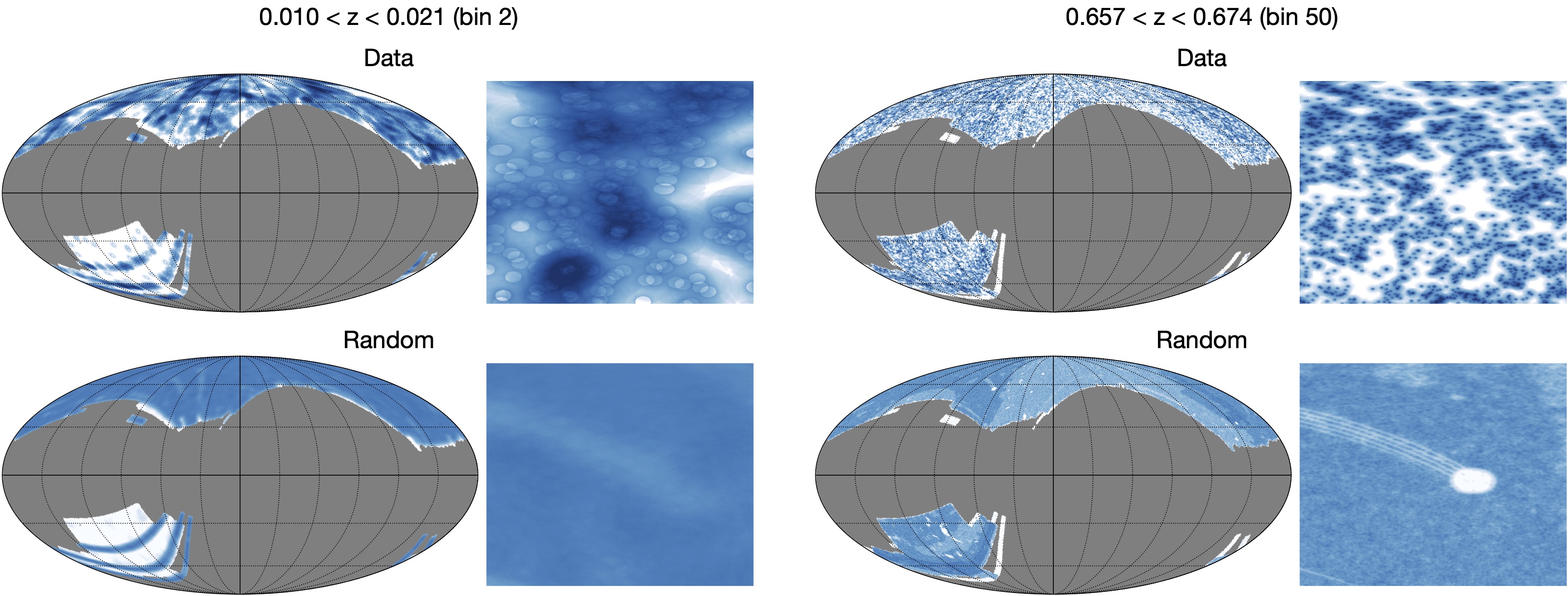}
\caption{Example activation maps for two reference redshift slices, $0.010<z<0.021$ (left) and $0.657<z<0.674$ (right). For each slice, we show the data (top) and random (bottom) activation maps together with a zoom-in. Each reference object activates pixels within a fixed range of projected physical separations $\rp$, weighted by $W'(\theta)$. At low redshift, the large angular scale produces broad, overlapping annuli, whereas at higher redshift the annuli become compact and sparsely distributed, closely tracing the underlying LSS. The random activation maps encode the survey window or selection function, including features such as bright-star masks (visible as a ringed hole in the lower-right zoom), whose effects are removed by the correlation estimators.}
    \label{fig:activation_maps}
\end{figure*}

\subsection{Pair-less Correlations with Activation Maps} \label{subsec:algorithm}

Conventionally, angular correlation functions are measured by calculating the separations of all cross-sample pairs and accumulating the resulting pair counts in angular bins. Brute-force computation scales as $\mathcal{O}(N^2)$ with the sample size $N$, while tree-based neighbor-search algorithms such as $k$-d trees \citep{kdtree} reduce the cost to $\mathcal{O}(N\log N)$ \citep{kdtree_complexity}. Nevertheless, evaluating correlations for samples containing $10^4$--$10^8$ objects in fine redshift bins remains expensive. Moreover, every new test sample requires repeating the full pair-counting calculation.

\tomographer\ instead performs the pair-finding step only once for the fixed reference sample, storing the result as a set of ``activation maps'' that encode the spatial pair information. The key to making these activation maps reusable for any new test sample is an intermediate pixelization layer, onto which any generalized test field $S(\phi)$, whether a source catalog or an intensity map, is projected. This representation converts all inputs into a common vector of pixel values that can be processed identically, while preserving the fixed pixel coordinates whose spatial relationships to the reference sample can be precomputed. As commonly adopted in the community, \tomographer\ uses the HEALPix pixelization in Galactic coordinates with $\mathrm{NSIDE}=2048$ resolution, corresponding to $50{,}331{,}648$ pixels of approximately $1.7\arcmin$ across, with an effective Gaussian full width at half-maximum (FWHM) of $1.24\arcmin$. This is sufficient to resolve the smallest clustering scale used here, $0.5\,h^{-1}$ physical Mpc, at all redshifts (see Section~\ref{subsec:scales}). In principle, however, the activation-map approach is compatible with any pixelization scheme.

To prepare for activation-map generation, the spectroscopic reference sample is retained at its full angular resolution without pixelization. In \tomographer, it is divided into logarithmically spaced $1+z$ bins using either a fiducial 160 bin configuration over $0<z<4.2$ for science analyses or a coarse 40 bin configuration for fast exploration (Section~\ref{subsec:config}).

We now build the activation maps, one for the data and one for the random catalog in each reference redshift bin $z_i$, on the full HEALPix grid. For each pixel centered at position $\phi$, we identify neighboring reference objects, indexed by $j$, within that redshift slice using a tree-based algorithm to search out to $\theta_{\rm max}$, then retain those within the annulus $\theta_{\rm min}<\theta_j<\theta_{\rm max}$ given the adopted physical separation (Section~\ref{subsec:scales}). To compute the activation value of the central pixel, the weighted pair counts, with per-pair weight $W'(\theta_j)$, are summed. Repeating this calculation over all pixels produces the full-sky data and random activation maps, which we denote in vector form by $\mathbf{A}_i^D$ and $\mathbf{A}_i^R$. Their components, or pixel values, $A_i^D(\phi)$ and $A_i^R(\phi)$ are given by
\begin{eqnarray}
A_i^{D}(\phi)
&\hspace{1em}\propto&
\sum_{j\in D_r(z_i),\,\theta_{\rm min}<\theta_j<\theta_{\rm max}}
W'(\theta_j),
\nonumber\\
A_i^{R}(\phi)
&\hspace{1em}\propto&
\sum_{j\in R_r(z_i),\,\theta_{\rm min}<\theta_j<\theta_{\rm max}}
W'(\theta_j).
\label{eq:activation_maps}
\end{eqnarray}
We adopt a per-pair weight of $W'(\theta)=\theta^{\gamma-1}$, which differs from the correlation function weight $W(\theta)=\theta^\gamma$ in Equation~(\ref{eq:weighting}) by a factor of $\theta$ to account for the surface element. As shown in Appendix~\ref{app:weighting}, the two formulations are equivalent up to a known multiplicative factor. The activation maps $\mathbf{A}_i$ may be stored with arbitrary normalization. At runtime, they are multiplied by the user-supplied weight field and survey footprint, then normalized to unit total weight. The resulting $\mathbf{A}_i$ form a set of basis templates associated with the LSS in independent reference redshift slices, onto which the test field can be projected to isolate its response at each redshift.

Figure~\ref{fig:activation_maps} shows examples of activation maps for two reference redshift slices. At low redshift, the larger angular size of the fixed physical aperture produces clearly visible annuli around individual reference galaxies, whereas at higher redshift the smaller annuli give the maps a more fine-grained appearance, with both tracing the underlying LSS. The corresponding random activation maps in the lower row encode the survey window of the reference sample, including the footprint, masked regions, and spatial selection function, allowing these observational effects to be corrected.

Having constructed the activation maps, we now show how they are used to evaluate the clustering estimators at runtime. To evaluate the estimators in Section~\ref{subsec:estimator}, the test observable $\Delta S(\phi)$ and its corresponding random field $\Delta S^R(\phi)$, if available, together with the reference density field in each redshift slice, $N_r(\phi,z_i)$, are projected onto the same HEALPix grid as the activation maps and represented in vector form as $\mathbf{\Delta S}$, $\mathbf{\Delta S}^R$, and $\mathbf{N}_{r,i}$, respectively. While the activation maps contain all the information needed to evaluate the cross correlations $\bar{w}_{tr}$, the reference maps $\mathbf{N}_{r,i}$ are retained at runtime to recompute the reference autocorrelation $\bar{w}_{rr}$, and hence the reference bias $b_r$, for the run-specific survey footprint, thereby accounting for changes in the effective mixture of reference subsamples.

The generalized test--reference pair accumulations for our Landy--Szalay-like estimator (Equation~(\ref{eq:w_tr_ls})), or the corresponding subset for the Davis--Peebles-like estimator (Equation~(\ref{eq:w_tr_dp})) when no test random catalog is available, reduce to
\begin{eqnarray}
\makebox[0.48\linewidth][l]{$
D_tD_r(z_i)
=
\mathbf{\Delta S}\cdot\mathbf{A}_i^D,
$}
&&
D_tR_r(z_i)
=
\mathbf{\Delta S}\cdot\mathbf{A}_i^R,
\nonumber\\
\makebox[0.48\linewidth][l]{$
R_tD_r(z_i)
=
\mathbf{\Delta S}^{R}\cdot\mathbf{A}_i^D,
$}
&&
R_tR_r(z_i)
=
\mathbf{\Delta S}^{R}\cdot\mathbf{A}_i^R,\nonumber\\
\label{eq:pair_dot}
\end{eqnarray}
where the dot denotes the inner product over HEALPix pixels within the survey footprint. The reference autocorrelation terms in the standard Davis--Peebles estimator (Equation~(\ref{eq:w_rr_dp})) are evaluated analogously,
\begin{eqnarray}
D_rD_r(z_i)
&=&
\mathbf{N}_{r,i}\cdot\mathbf{A}_i^D,
\nonumber\\
D_rR_r(z_i)
&=&
\mathbf{N}_{r,i}\cdot\mathbf{A}_i^R,
\label{eq:pair_dot_rr}
\end{eqnarray}
yielding $\bar{w}_{rr}$ and hence the reference bias $b_r$ through Equation~(\ref{eq:b_r}).

Thus, all correlation measurements reduce to inner products that represent the projection amplitudes of pixelized fields onto the redshift-dependent activation basis, eliminating the need to explicitly evaluate pair separations for each new test sample. The unit-weight normalization of $\mathbf{A}_i$ preserves the native units and scales of the projected test observables $S$.

In practice, these inner products are computed using highly optimized vectorized operations in Python, making the runtime algorithm effectively ``pair-less'' and reducing the cost of each correlation measurement to $\mathcal{O}(1)$, independent of the sizes of both the test and reference samples. More specifically, to accelerate the computation, we use \texttt{NumExpr} \citep{2018zndo...1492916C}, an efficient array expression evaluator, to perform the inner-product calculations.
A full end-to-end clustering-redshift analysis over 160 redshift bins typically completes in minutes, and sometimes seconds, on a single workstation, with runtime dominated by map I/O, optional foreground mitigation (filtering and/or template regression), and resampling for error estimation rather than the correlation calculation itself.

\begin{deluxetable*}{ccccccc}
\tablehead{\colhead{Sample} & \colhead{Area$^a$} & \colhead{N$^b$} & \colhead{N/Area} & \colhead{$z_{\rm med}$$^c$} & \colhead{$[z_{5}, z_{95}]$$^d$} & \colhead{References}\\ \colhead{ } & \colhead{$\rm deg^2$} & \colhead{ } & \colhead{$\rm deg^{-2}$} & \colhead{ } & \colhead{ } & \colhead{ }}
\startdata
SDSS MAIN & 7660 & 550,141 & 71.8 & 0.09 & [0.03, 0.19] & (1, 2, 3) \\
BOSS LOWZ & 9150 & 463,044 & 50.6 & 0.31 & [0.09, 0.44] & (4) \\
BOSS CMASS & 9960 & 849,637 & 85.3 & 0.54 & [0.43, 0.69] & (4) \\
eBOSS LRG & 5090 & 174,816 & 34.3 & 0.74 & [0.62, 0.91] & (5) \\
eBOSS ELG & 1100 & 173,736 & 157.8 & 0.83 & [0.69, 1.03] & (6) \\
eBOSS QSO & 5040 & 343,708 & 68.2 & 1.51 & [0.90, 2.11] & (5) \\
BOSS QSO CORE & 9670 & 97,675 & 10.1 & 2.47 & [2.20, 3.23] & (7, 8) \\
BOSS QSO HIGHZ & 9670 & 5612 & 0.6 & 3.7 & [3.45, 4.11] & (9) \\
non-LSS QSO & 10,440 & 416,571 & 39.9 & 1.48 & [0.33, 3.33] & (10, 11) \\ \hline
Total & 10,440 & 3,074,940 & 294.5$^e$ & 0.55$^e$ & [0.07, 2.42]$^e$ &
\enddata
\vspace{0.4em}
\textbf{Notes.} $^a$ Effective unmasked sky area. $^b$ Number of sources. $^c$ Median redshift. $^d$ 5th and 95th percentile redshifts. $^e$ Spatially varying.
\vspace{0.4em}
\textbf{References.} (1) \citet{2002AJ....124.1810S}, (2) \citet{2005AJ....129.2562B}, (3) \citet{2015MNRAS.449..835R}, (4) \citet{2016MNRAS.455.1553R}, (5) \citet{2020MNRAS.498.2354R}, (6) \citet{2021MNRAS.500.3254R}, (7) \citet{2012MNRAS.424..933W}, (8)
\citet{2015MNRAS.453.2779E}, (9)
\citet{paris2017}, (10)
\citet{2010AJ....139.2360S}, (11) \citet{2020ApJS..250....8L}.
\caption{SDSS Spectroscopic Redshift Reference}
\label{tab:reference}
\end{deluxetable*}

\vspace{-1\baselineskip}
\subsection{Reference Sample} \label{subsec:reference}

The statistical power of a clustering-redshift measurement is set by the reference sample. In the current release of \tomographer, the built-in reference sample consists of nine legacy SDSS spectroscopic LSS subsamples, summarized in Table~\ref{tab:reference}, with their sky coverage shown in Appendix~\ref{app:ref_mix}. Together they contain 3,074,940 spectroscopically confirmed galaxies and quasars over an effective area of $10{,}440\,{\rm deg}^2$, providing continuous redshift coverage over $0<z\lesssim4.2$.

\begin{figure*}[!t]
    \centering
    \includegraphics[width=0.975\textwidth]{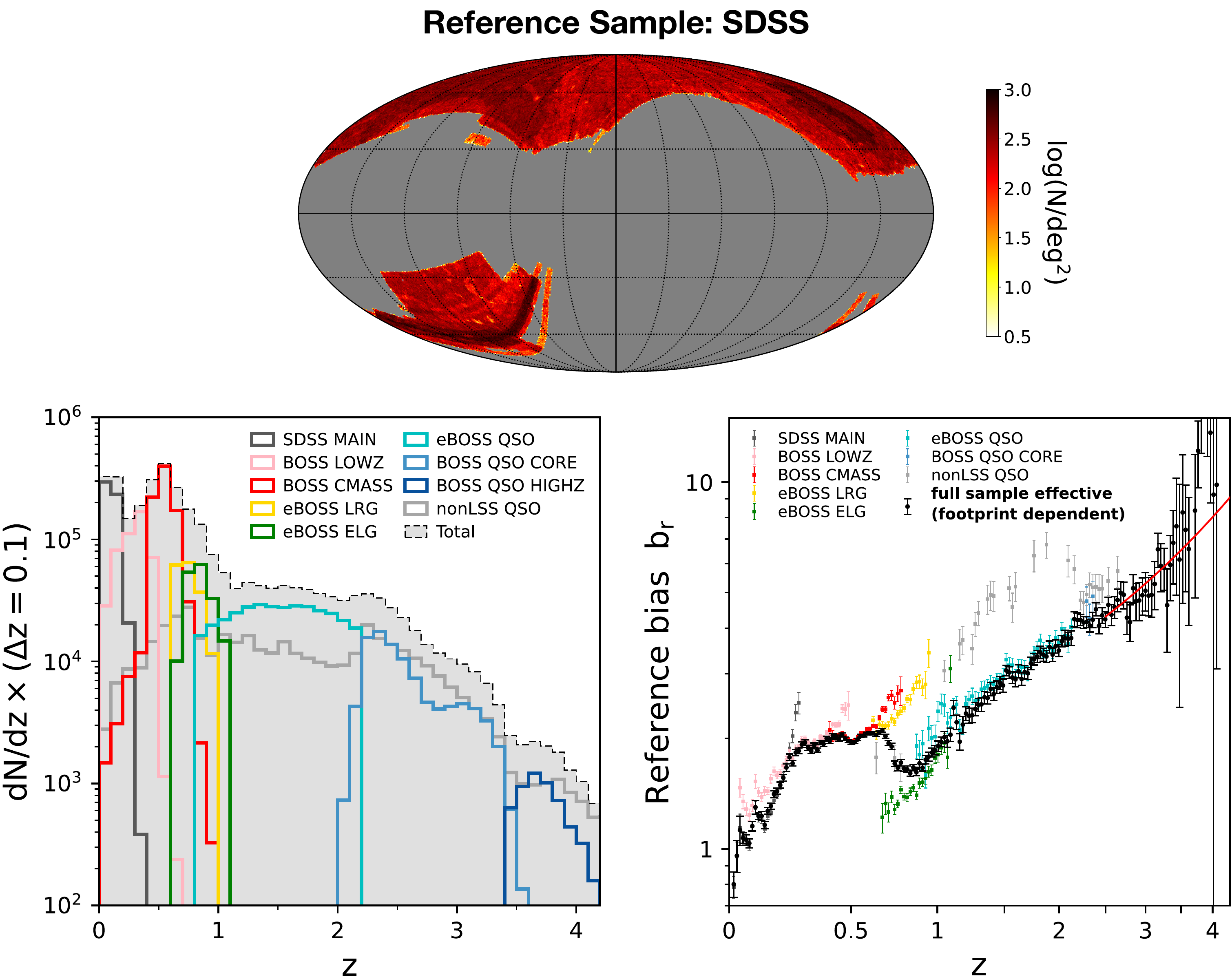}
\caption{The SDSS spectroscopic reference sample. Top: surface density of the combined sample over its $10{,}440\,{\rm deg}^2$ footprint. Bottom left: redshift distributions of the nine subsamples (Table~\ref{tab:reference}) and their combination, providing continuous coverage over $0<z\lesssim4$. Bottom right: reference bias $b_r(z)$ measured from the autocorrelation of each redshift slice, with the effective bias of the combined sample in black. At $z\gtrsim2.5$, a smooth parametric fit is used in place of the noisier measurements, with negligible impact on the final \tomographer\ uncertainties validated in Section~\ref{subsec:val_specz}. The lower panels were previously presented in \citet{chiang25}.}
\label{fig:ref_N_of_z_b_of_z}
\end{figure*}

At $z\lesssim1$ the reference is dominated by galaxy samples---SDSS MAIN \citep{2002AJ....124.1810S, 2005AJ....129.2562B, 2015MNRAS.449..835R}, the Baryon Oscillation Spectroscopic Survey (BOSS) LOWZ and CMASS samples \citep{2016MNRAS.455.1553R}, and extended BOSS (eBOSS) luminous red galaxies (LRGs) and emission-line galaxies \citep[ELGs;][]{2020MNRAS.498.2354R, 2021MNRAS.500.3254R}---and at higher redshifts by quasars (QSOs) from eBOSS \citep{2020MNRAS.498.2354R}, the BOSS CORE and high-$z$ samples \citep{2012MNRAS.424..933W, 2015MNRAS.453.2779E, paris2017}, and a ``non-LSS'' compilation of quasars from \citet{2010AJ....139.2360S} and \citet{2020ApJS..250....8L} outside the homogeneously selected LSS set. For samples drawn from the SDSS LSS catalogs, which provide a well-characterized selection function through accompanying random catalogs and weights, we use these products directly. For the non-LSS quasars, we instead construct an approximate random catalog from the SDSS imaging footprint and masks to reproduce the selection function. 

Table~\ref{tab:reference} summarizes the key properties of the reference subsamples, including their footprints, surface densities, and redshift ranges, while Figure~\ref{fig:ref_N_of_z_b_of_z} shows their combined sky density, redshift distributions, and measured clustering biases. The redshift distributions and clustering biases were previously presented in \citet{chiang25} and are reproduced here for completeness. The effective bias of the full reference sample increases with redshift, exhibiting an intermediate-redshift plateau driven by LRGs before transitioning to ELGs and quasars. Sky maps and footprints of the individual subsamples are provided in Appendix~\ref{app:ref_mix}.

\begin{figure*}[t]
    \begin{center}
         \includegraphics[width=0.88\textwidth]{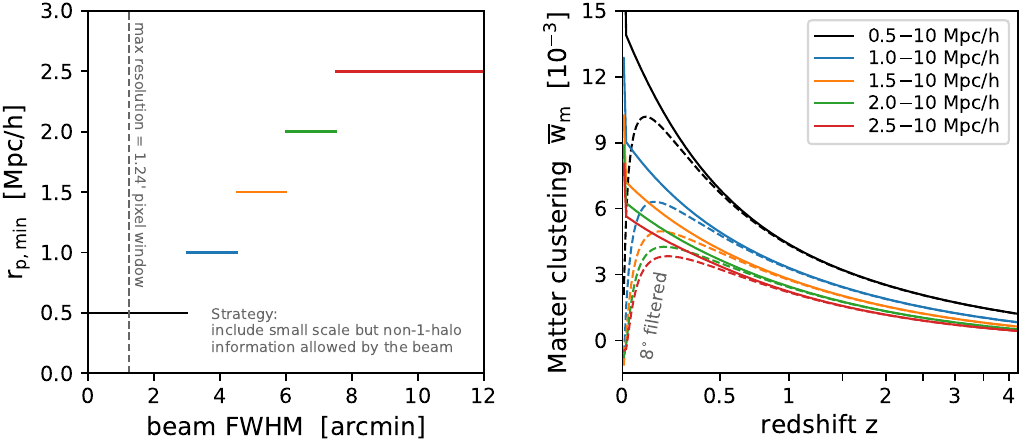}
    \end{center}
    \caption{Scale choices and the matter-clustering kernel. Left: the minimum projected separation $r_{\rm p,min}$ used in the correlation measurements as a function of the beam FWHM of the test map. The strategy is to include the small-scale information resolved by the beam while excluding scales dominated by the one-halo term and the instrument response; the vertical dashed line marks the $\mathrm{NSIDE}=2048$ pixel window. Right: the integrated matter-clustering amplitude $\wbar_{\rm m}(z)$ for each $\rp$ range (solid; $r_{\rm p,max}=10\,h^{-1}$Mpc throughout), used in Equation~(\ref{eq:master}). Dashed curves show the effect of the $8\degr$ high-pass filter, one of several built-in filtering options for foreground mitigation and suppression of large-scale systematics.}
    \label{fig:w_m}
\end{figure*}

The nine reference subsamples are combined into a single reference catalog, with their random catalogs first trimmed to a common random-to-data ratio before being merged accordingly. Although the resulting reference sample is not itself homogeneous, and its mixture and effective bias evolution depend on the survey footprint, this does not bias clustering redshifts. As discussed in Section~\ref{subsec:projected}, the observable $S(\phi)$ is required to have a unique $b(z)\,\mathrm{d}S/\mathrm{d}z(z)$, while each homogeneous reference subsample provides an independent estimate of the same quantity, which we demonstrate using a specific example in Section~\ref{subsec:val_bref}. Because the estimator (Equation~(\ref{eq:master})) is linear, these estimates can be combined. \tomographer\ therefore performs the combination at the catalog level before activation-map generation and runtime summary-statistic measurements, substantially reducing the computational cost compared with treating the reference subsamples separately. 

As described above, the reference bias is measured directly from the reference sample within the survey footprint of each run. Only when the SNR of $\bar{w}_{rr}(z)$ is insufficient for a reliable measurement (e.g., for a small survey footprint) does \tomographer\ fall back to a precomputed effective reference bias derived from the full reference catalog.

\subsection{Clustering Scales and Matched Matter Anchors} \label{subsec:scales}

The angular range $(\theta_{\rm min},\theta_{\rm max})$ used to define the scale-integrated clustering amplitude (Equation~(\ref{eq:weighting})) is somewhat tunable but is applied consistently to all correlation measurements, $\wbar_{tr}$, $\wbar_{rr}$, and the corresponding theoretical matter anchor of $\wbar_{\rm m}$ calculation. We set $(\theta_{\rm min},\theta_{\rm max})$ through a fixed range of projected physical separations, $(r_{\rm p,min}, r_{\rm p,max})$, evaluated at the redshift of each reference slice. We use physical, rather than comoving, separations for a weaker redshift evolution of the corresponding angular scale and a more physical inner cutoff relative to the sizes of nonlinear structures and halos decoupled from the Hubble flow. We adopt $r_{\rm p,max}=10\,h^{-1}$~Mpc, together with a maximum angular radius of $15^\circ$ (affecting only the lowest-redshift bins), to limit sensitivity to wide-angle systematics, such as residual Galactic extinction and spatial variations in photometric calibration \citep{2011MNRAS.417.1350R,2013MNRAS.432.2945H}.

The inner scale cut is chosen to suppress one-halo contributions, where galaxy clustering is no longer governed primarily by gravity but by strongly sample-dependent astrophysical effects of galaxy formation, which could cause Equation~(\ref{eq:master}) to break down. Naturally, $r_{\rm p,min}$ must also exceed both the pixel scale ($1.24\arcmin$ Gaussian FWHM for $\mathrm{NSIDE}=2048$) and the instrumental beam. We adopt a default $r_{\rm p,min}=0.5\,h^{-1}$Mpc for source catalogs whose astrometric precision exceeds the pixel resolution and for high-resolution intensity maps, retaining most of the small-scale signal while largely excluding one-halo contributions. Because each $(r_{\rm p,min},r_{\rm p,max})$ choice requires a separate set of precomputed activation maps, supporting additional scale ranges substantially increases the distributed data volume (Section~\ref{subsec:config}). We therefore make available five configurations, which \tomographer\ automatically selects according to the beam FWHM $\theta_{\rm FWHM}$: $r_{\rm p,min}=0.5$, $1.0$, $1.5$, $2.0$, and $2.5\,h^{-1}$Mpc for $\theta_{\rm FWHM}<3.0'$, $3.0'\le\theta_{\rm FWHM}<4.5'$, $4.5'\le\theta_{\rm FWHM}<6.0'$, $6.0'\le\theta_{\rm FWHM}<7.5'$, and $\theta_{\rm FWHM}\ge7.5'$, respectively. Figure~\ref{fig:w_m} (left) summarizes the resulting scale-selection strategy as a function of beam FWHM.

The corresponding matter-clustering amplitude, $\wbar_{\rm m}(z)$, calculated using Equations~(\ref{eq:wm}) and (\ref{eq:weighting}) and shown in Figure~\ref{fig:w_m} (right), is determined by both the adopted clustering scale range and the assumed cosmology. It is computed from the nonlinear matter power spectrum $P_{\rm m}^{\rm NL}$ for our assumed cosmology, using \texttt{CLASS} \citep{2011JCAP...07..034B} with the \texttt{halofit} prescription \citep{2003MNRAS.341.1311S}. The $\mathrm{NSIDE}=2048$ pixel window is also included for completeness, although its effect is negligible because even the smallest adopted $r_{p,\min}$ exceeds the effective pixel scale. A separate $\wbar_{\rm m}(z)$ is calculated for each precomputed scale configuration and enters the redshift inference in Equation~(\ref{eq:master}) as a correction term.

\tomographer\ optionally supports foreground mitigation through spatial filtering of the input maps. Because filtering modifies the measured clustering signal, the same operation is applied when computing $\wbar_{\rm m}(z)$. The resulting curves for one example filtering scale are shown as dashed lines in Figure~\ref{fig:w_m} (right), differing primarily at low redshift, where the fixed physical aperture subtends the largest angular scales.

\subsection{Analysis Configuration and Considerations} \label{subsec:config}

The primary user interface of \tomographer\ is a configuration file (by default, \texttt{conf.ini}), which specifies the analysis settings together with the locations of the required input files. These may include the test sample, survey footprint or random catalog, optional masking and weighting maps, and other auxiliary inputs depending on the analysis. Below, we describe the principal analysis options together with the considerations for choosing appropriate settings for different source catalogs and intensity maps, while also highlighting the underlying design principles we adopted.

\emph{Redshift binning and angular scales.}
\tomographer\ provides two activation-map configurations. The \emph{quick} mode uses 40 redshift bins over a fixed range of $\rp=0.5$--$10\,h^{-1}$Mpc; these data products are included with the package and are intended for testing and quick-turnaround exploration of source catalogs and high-resolution (beam FWHM $\lesssim3\arcmin$) intensity maps. The \emph{fiducial} mode uses the full 160 bin activation maps, with the clustering scale $\rp$ selected automatically from the beam FWHM following Section~\ref{subsec:scales}. The corresponding activation maps are downloaded on demand from Zenodo \citep[][see Appendix~\ref{app:built_in_masks}]{chiang_2026_20155554}.\footnote{\url{https://zenodo.org/records/20155554}} This is the recommended configuration for science-grade analyses. Users may optionally rebin the finely sampled fiducial results to improve the per-bin SNR.

\emph{Footprint, masking, and spatial weighting.}
Robust clustering measurements rely critically on an accurate characterization of the test-sample selection function, which is inherently sample dependent. Internally, \tomographer\ represents survey footprints, binary masks, and continuous weighting maps using a unified spatial weighting framework. Pixels with zero weight are excluded from the analysis, while intermediate weights proportionally reduce the contribution of the corresponding test-sample pixels to the effective pair counts. All supplied footprint, mask, and weighting maps are combined multiplicatively into a single effective weight field.

The survey footprint may be specified in one of three ways. In the recommended \texttt{user\_defined} mode, users provide either a random catalog or a footprint mask ($\mathrm{NSIDE}=2048$ HEALPix map in Galactic coordinates) describing the survey selection function. Alternatively, users may select the \texttt{full\_sky} mode to analyze all pixels overlapping the reference footprint, or the \texttt{auto\_detection} mode, which estimates a coarse survey footprint directly from the data.\footnote{This procedure iteratively degrades the source-map resolution until a stopping criterion is met, with nonzero pixels defining the survey footprint.} Although convenient when no footprint information is available, the latter provides only an approximate footprint and may be subject to boundary effects.

We provide several built-in masks for common applications, summarized in Appendix~\ref{app:built_in_masks}. Users may additionally supply arbitrary $\mathrm{NSIDE}=2048$ HEALPix masking or weighting maps.

\emph{Foreground mitigation.}
Depending on the characteristics of the input map, users may choose to reduce Galactic foreground contamination through template regression. \tomographer\ provides two built-in foreground templates, summarized in Appendix~\ref{app:built_in_masks}, which can be enabled directly in the configuration. Cleaning with other custom templates may instead be applied by users prior to running the pipeline. High-pass filtering described in Section~\ref{subsec:scales} may additionally be enabled, with predefined filter scales of $2\degr$, $4\degr$, $8\degr$, or $16\degr$, or no filtering. Because high-pass filtering modifies the clustering signal, its effect is included when computing $\wbar_{\rm m}(z)$; no such correction is needed for template regression. The two operations may be enabled independently or together.

\subsection{Bootstrap Uncertainty Estimation} \label{subsec:errors}

Statistical uncertainties in \tomographer\ are estimated using the spatial block-bootstrap procedure of \citet{2008ApJ...681..726L}, following the implementation of \citet{chiang23}. The sky is partitioned into $\mathrm{NSIDE}=16$ HEALPix regions, substantially coarser than the pixel scale, which serve as approximately independent bootstrap units. Bootstrap realizations are generated by resampling these regions with replacement and propagated through the full clustering-redshift estimator. All realizations are evaluated simultaneously in a vectorized implementation, producing an ensemble of reconstructed $b\,({\rm d}N/{\rm d}z)$ or $b\,({\rm d}I/{\rm d}z)$ that provides the statistical uncertainties and covariance between redshift bins.

Rather than generating bootstrap realizations during each run, \tomographer\ distributes a library of bootstrap pixel-ID lists drawn once in advance as part of the data products. This ensures that independent \tomographer\ analyses automatically share a common set of bootstrap realizations, enabling covariance estimation across different runs for multiband tomography \citep[e.g.,][]{chiang25}. The bootstrap indexing maps are incorporated into the same map-inner-product framework, effectively augmenting and modulating the activation maps (Equations~(\ref{eq:pair_dot}) and (\ref{eq:pair_dot_rr})), introducing significant but not prohibitive computational overhead.

The number of bootstrap realizations is configurable. Increasing the number of realizations improves the precision of the estimated covariance at the cost of additional runtime and memory usage. We recommend using $10$--$30$ realizations for testing and at least $50$ for science analyses.

The statistical accuracy of this bootstrap procedure is validated in Section~\ref{subsec:val_specz}, where the resulting error bars are shown to provide unbiased uncertainty estimates relative to known truth distributions.

\begin{figure*}[!t]
    \begin{center}
         \includegraphics[width=0.75\textwidth]{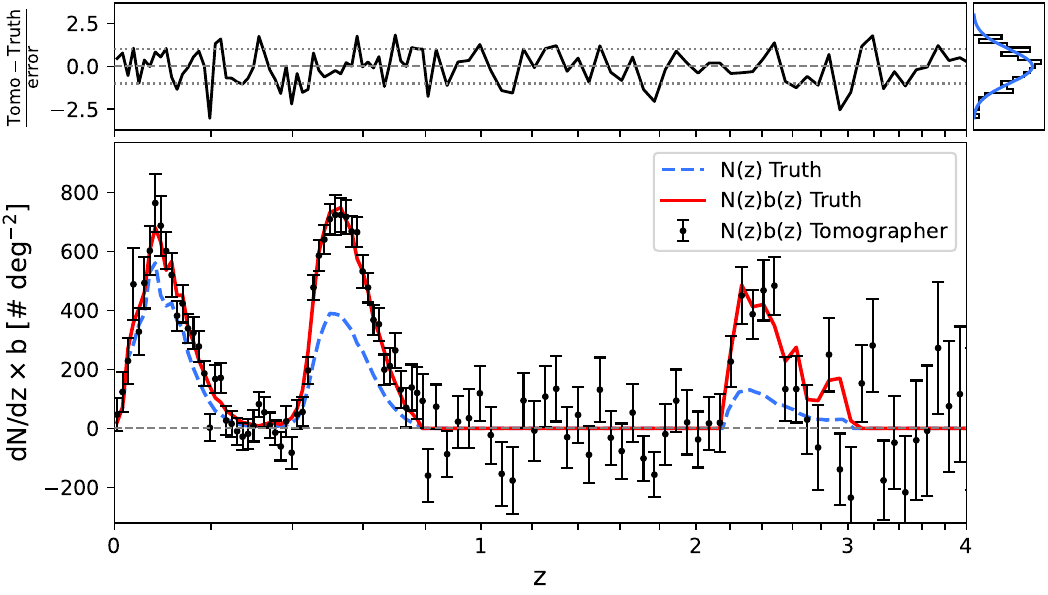}
    \end{center}
    \vspace{0.em}
    \begin{center}
         \includegraphics[width=0.75\textwidth]{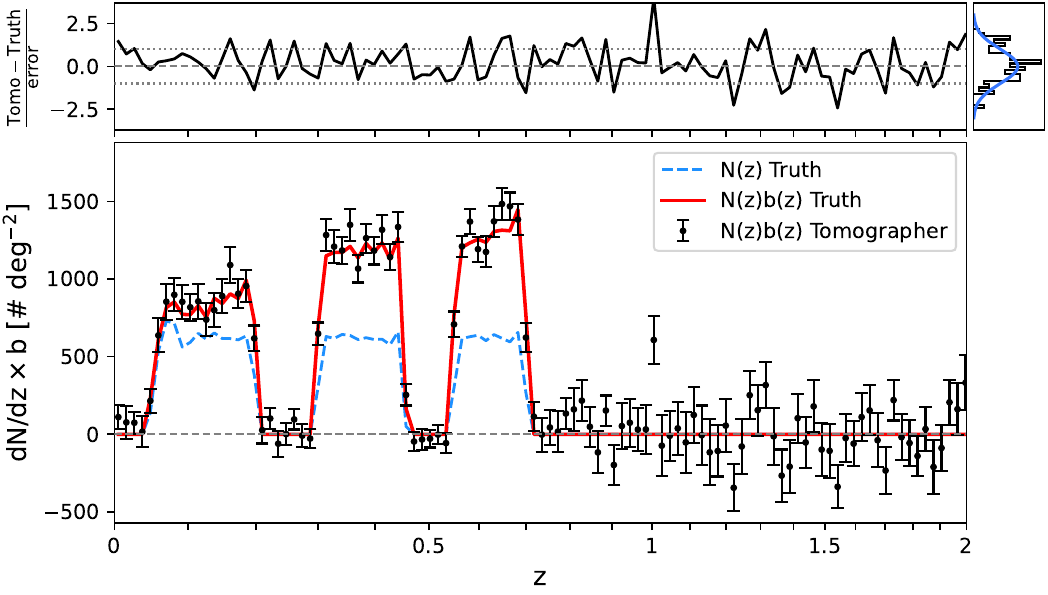}
    \end{center}
\caption{Validation against samples with known spectroscopic redshift distributions. Top: a composite test sample built from SDSS MAIN galaxies, BOSS CMASS galaxies, and BOSS CORE QSOs, with the QSO sample duplicated by a factor of 10 for comparable normalization. Bottom: MAIN, LOWZ, and CMASS galaxies trimmed into three sharp top-hat redshift intervals. In each case, black points show the \tomographer\ result, while red and blue dashed curves show the true $b\,({\rm d}N/{\rm d}z)$ and ${\rm d}N/{\rm d}z$, respectively, provided or measured from the spectroscopic catalog. The upper strips show \tomographer\ residuals normalized by the estimated uncertainties, with side panels comparing their distribution to a unit Gaussian (solid curve). The recovered distributions are unbiased, resolve sharp features without ringing, show no spurious signal outside the true redshift range, and yield largely unbiased error bars.}
\label{fig:general_test_wide}
\end{figure*}

\begin{figure*}[h!]
    \begin{center}
         \includegraphics[width=0.87\textwidth]{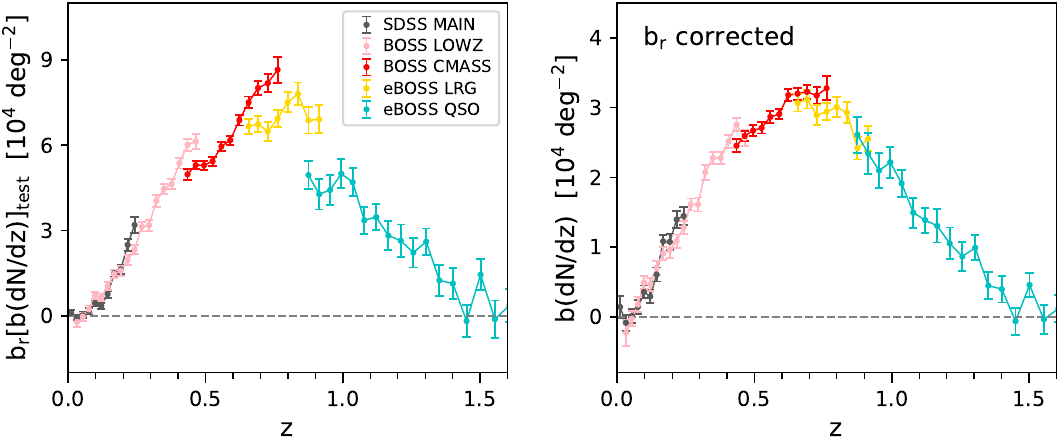}
    \end{center}
\caption{Reference-bias correction, demonstrated using a customized in-house version of \tomographer. A single broad test sample of DESI Legacy Imaging Surveys sources with $m_z<22.5$ is analyzed against five different reference subsamples. Left: the raw clustering-redshift measurements, corrected only for $\bar{w}_m$ and therefore proportional to $b_r\,b_t\,({\rm d}N_t/{\rm d}z)$, differ in amplitude because of the distinct clustering biases of the reference tracers. Right: after applying the measured $b_r(z)$ correction (Equation~(\ref{eq:master})), all five measurements collapse onto a consistent track of $b\,({\rm d}N/{\rm d}z)$, demonstrating independence from the choice of reference population.}
    \label{fig:b_ref_correction}
\end{figure*}

\begin{figure*}[!t]
    \begin{center}
         \includegraphics[width=0.85\textwidth]{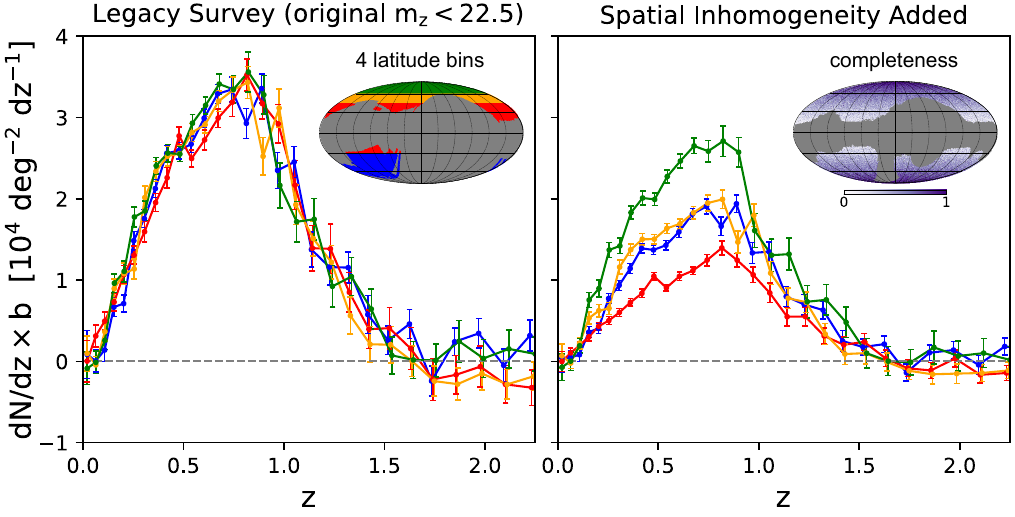}
    \end{center}
\caption{Selection diagnostics and homogeneity tests using Legacy Surveys galaxies selected to $m_z<22.5$. Left: the test sample is split into four equal-area Galactic-latitude zones (inset) and analyzed independently. The distributions recovered by \tomographer\ agree within the uncertainties, demonstrating homogeneity and robustness to variations in survey depth, extinction, and stellar density. Right: a synthetic angular selection function is imposed through a large-scale completeness gradient (inset) and spatially dependent downsampling. \tomographer\ recovers the correct \emph{shape}, while the amplitudes follow the imposed completeness, showing that angular completeness affects only the normalization of ${\rm d}N/{\rm d}z$.}
    \label{fig:glat_splits}
\end{figure*}

\section{Validation}\label{sec:validation}

We validate the bias-weighted redshift distributions produced by \tomographer\ with SDSS spectroscopic references through a series of controlled tests spanning a variety of input datasets, including SDSS, the Legacy Surveys \citep{2019AJ....157..168D}, and unWISE \citep{2019ApJS..240...30S}. These tests cover known redshift recovery, independence from the choice of reference subsample, synthetic intensity maps generated from resolved sources, and comparisons to independent external measurements.

\subsection{Recovery of Known Spectroscopic Distributions} \label{subsec:val_specz}

Our first set of validations uses test samples drawn from the SDSS LSS catalogs, for which the true ${\rm d}N/{\rm d}z(z)$ is known spectroscopically, and $b(z)$ is measured from their autocorrelations, providing the true $b(z)\,{\rm d}N/{\rm d}z(z)$ for comparison. Although these samples partially overlap with the built-in reference catalog in \tomographer, the overlap does not affect the validation because self-pairs fall below $\theta_{\rm min}$ and therefore never enter the clustering-redshift measurement.

Figure~\ref{fig:general_test_wide} (top) presents a \emph{general} case using a composite test sample consisting of MAIN galaxies at $z\sim0.1$, CMASS galaxies at $z\sim0.55$, and BOSS CORE QSOs at $z\sim2.5$. For better visualization, the QSO sample is duplicated by a factor of 10 so that its overall normalization is comparable to that of the two lower-redshift galaxy samples. \tomographer\ accurately recovers all three peaks in $b(z)\,{\rm d}N/{\rm d}z(z)$ while remaining consistent with zero outside the relevant redshift spans. Comparing the true ${\rm d}N/{\rm d}z(z)$ and $b(z)\,{\rm d}N/{\rm d}z(z)$ curves also illustrates the effect of bias weighting: the more highly biased CMASS and QSO populations produce stronger peaks relative to their raw number counts.

Figure~\ref{fig:general_test_wide} (bottom) presents a \emph{spiky} case designed to test the recovery of sharp features in redshift distributions. We construct a test sample by controlled downsampling of the parent MAIN, LOWZ, and CMASS spectroscopic catalogs, producing three top-hat redshift distributions spanning $0.07\lesssim z\lesssim0.27$, $0.30\lesssim z\lesssim0.47$, and $0.55\lesssim z\lesssim0.70$. \tomographer\ accurately reproduces all the sharp edges, with no evidence of smoothing or ringing, while the gaps between the top-hat distributions remain consistent with zero. By comparison, any photometric-redshift method applied without spectroscopic information would substantially smear these sharp features and introduce leakage beyond the true boundaries.

In both tests, the right-hand panels in Figure~\ref{fig:general_test_wide} show histograms of the $b\,({\rm d}N/{\rm d}z)$ residuals normalized by their estimated uncertainties, $({\rm Tomographer}-{\rm truth})/\sigma$. Both are consistent with a unit Gaussian (blue curves; not fitted), indicating that the bootstrap error bars in \tomographer\ are largely unbiased. This agreement also indicates that neglected calibration uncertainties, including those in the reference bias $b_r$, are small compared with the typical statistical errors.

\subsection{Independence of the Reference Sample} \label{subsec:val_bref}

A robust clustering-redshift estimate should not depend on the choice of reference population. Since the underlying matter density field is not directly observable, practical implementations necessarily rely on galaxy populations as biased tracers, each with its own clustering bias, redshift distribution, and selection function. Consistency across independent reference populations therefore provides a stringent internal validation of the clustering-redshift formalism.

We verify this by analyzing a single broad test sample of Legacy Surveys galaxies \citep{2019AJ....157..168D}, selected with an extinction-corrected $z$-band magnitude cut of $m_z<22.5$, independently against five SDSS reference subsamples spanning a wide range of tracer types (MAIN, LOWZ, CMASS, eBOSS LRG, and QSO). Figure~\ref{fig:b_ref_correction} first shows the raw measurements, proportional to $b_r\,b\,({\rm d}N/{\rm d}z)$, which exhibit clear amplitude differences reflecting the distinct clustering bias of each reference subsample. After applying the measured $b_r(z)$ correction in Equation~(\ref{eq:master}), the resulting test-sample $b\,({\rm d}N/{\rm d}z)$ estimates become mutually consistent within their uncertainties wherever they overlap. This validates the linear dependence on reference bias in Equation~(\ref{eq:master}) and supports the \tomographer\ approach of combining heterogeneous spectroscopic subsamples into a single reference catalog for improved SNR.

\subsection{Selection Diagnostics and Homogeneity} \label{subsec:val_footprint}

Real survey catalogs inevitably contain angular selection effects arising from depth variations, masking, dust extinction, stellar contamination, and other observational systematics. \tomographer\ can be used to test sample homogeneity empirically by repeating the clustering-redshift measurement over independent sky regions and comparing the recovered redshift distributions. Figure~\ref{fig:glat_splits} illustrates this approach using the same Legacy Surveys galaxy sample as in the previous test with $m_z<22.5$.

In the first case (left panel), we divide the survey footprint into four equal-area Galactic-latitude zones spanning a wide range of extinction and stellar density. The independently recovered redshift distributions agree within their uncertainties over the full redshift range, with no systematic trend with Galactic latitude. This directly supports that the dereddened $m_z<22.5$ sample satisfies the homogeneity criterion introduced in Section~\ref{subsec:projected}. Meanwhile, the results validate that \tomographer\ recovers consistent extragalactic redshift distributions despite substantial variations in sky conditions.

In the second exercise (right panel), we deliberately introduce a synthetic angular selection function by imposing a large-scale gradient in completeness and randomly downsampling galaxies accordingly. This represents a special but not uncommon case in which the selection function varies only spatially, rather than jointly with position and redshift. The measured distributions in \tomographer\ faithfully reproduce the imposed selection pattern across all four regions, demonstrating a valuable use of \tomographer\ for diagnosing such effects without prior knowledge of the selection function.

Together, these tests motivate using \tomographer\ over independent sky regions as a routine diagnostic of sample homogeneity and selection effects. If significant differences are found, the full-sample redshift distribution should be interpreted as an effective average rather than a unique global distribution, and any downstream cosmological or astrophysical interpretation should be treated with appropriate caution.

\begin{figure}[t!]
    \begin{center}
         \includegraphics[width=0.465\textwidth]{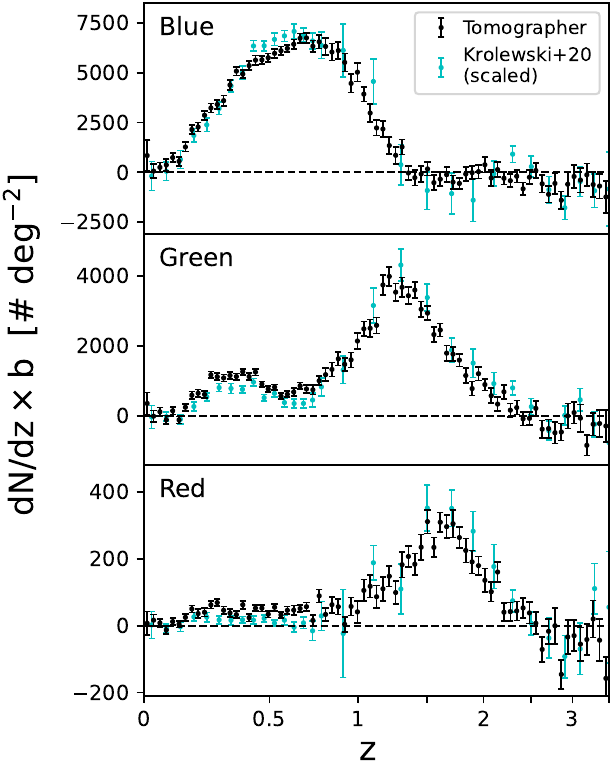}
    \end{center}
    \caption{Comparison with the clustering-redshift analysis of \citet{2020JCAP...05..047K}. The three unWISE galaxy samples, blue, green, and red, are analyzed with \tomographer\ (black) and compared to the results of \citet{2020JCAP...05..047K} (cyan), rescaled to match the integrated area of the physically normalized \tomographer\ measurements. The two analyses show good agreement in shape, peak location, and high-redshift cutoff, with minor differences in the detailed shapes. \tomographer\ additionally provides finer redshift sampling over a broader redshift range.}
    \label{fig:unWISE}
\end{figure}

\subsection{External Comparison: unWISE Galaxies} \label{subsec:val_unwise}

As an external end-to-end check, we compare \tomographer\ with the independent clustering-redshift analysis of the unWISE galaxy catalog by \citet{2020JCAP...05..047K}. They defined three color-selected samples of ``blue,'' ``green,'' and ``red'' galaxies with characteristic redshifts of $z\sim0.6$, 1.1, and 1.5, respectively, and measured their redshift responses by cross-correlating them with BOSS galaxies and eBOSS quasars using an independent analysis pipeline. Figure~\ref{fig:unWISE} compares the two sets of results, both in $b\,({\rm d}N/{\rm d}z)$. Since \tomographer\ directly recovers the absolute physical normalization of $b\,({\rm d}N/{\rm d}z)$, whereas the published distributions of \citet{2020JCAP...05..047K} were arbitrarily normalized to unit peak amplitude, we rescale the latter to have the same integrated area as ours for comparison. For all three samples, the overall recovered shapes, peak positions, and high-redshift cutoffs agree, with \tomographer\ providing slightly higher SNR, finer redshift resolution, and perhaps a better-behaved zero level (e.g., at $1.5<z<2$ for the blue sample). Some of these improvements can be attributed to the slightly larger reference sample used by \tomographer. Both analyses recover the bimodality of the green sample, revealing a lower-redshift galaxy population and a higher-redshift active galactic nucleus (AGN) component. Minor but noticeable differences remain, such as the peak height of the blue sample and the low-redshift tails of the green and red samples. These differences need not indicate localized discrepancies, but may instead reflect a different balance between the low- and high-redshift portions of the normalized distributions, arising from the different clustering scales adopted by the two analyses, any associated scale-dependent galaxy bias, and/or other minor implementation differences between the two pipelines.

\begin{figure*}[!t]
    \begin{center}
         \includegraphics[width=0.875\textwidth]{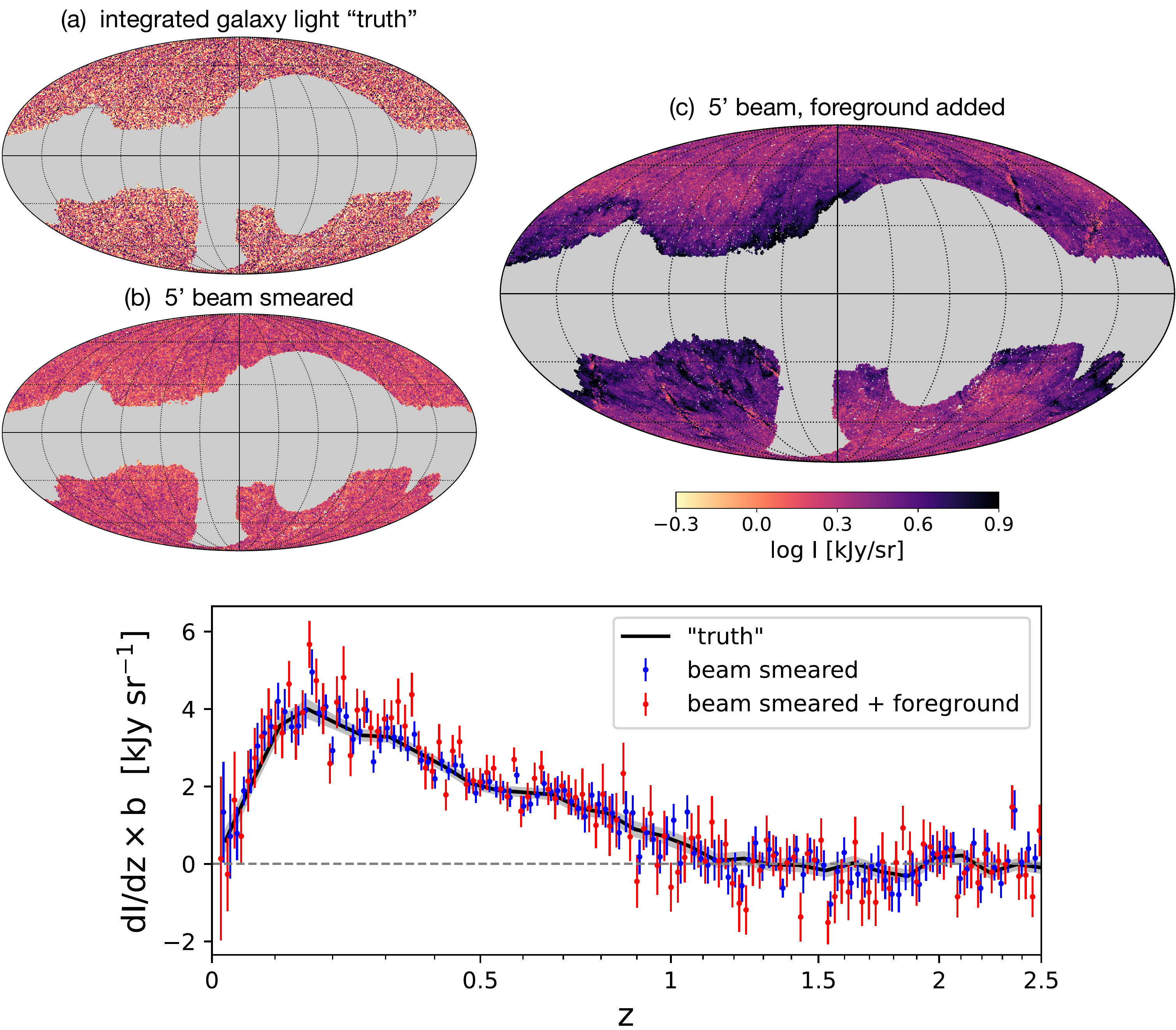}
    \end{center}
\caption{Validation of the intensity-mapping mode with a synthetic IGL map built from the integrated $z$-band light of Legacy Surveys galaxies down to magnitude 22.5. Top: the input maps---(a) the ``truth'' $z$-band IGL at native resolution, (b) the same map smoothed with a $5\arcmin$ beam, and (c) the smoothed map with a bright, spatially structured Galactic foreground constructed from WISE band-3 intensity tracing PAH emission in the interstellar medium \citep{2014ApJ...781....5M}. Bottom: the recovered $b\,({\rm d}I/{\rm d}z)$ by \tomographer\ for cases (b) and (c) (blue and red data points), compared to the reference (black curve with uncertainty band). Both measurements remain consistent with the high-resolution truth within the uncertainties, demonstrating robustness against beam smoothing and diffuse Galactic foregrounds.
}
    \label{fig:Syn_IM}
\end{figure*}

\subsection{Intensity-mapping Mode: Synthetic Map Test} \label{subsec:val_IM}

Having validated \tomographer\ with source catalogs, we next test its intensity-mapping mode using the same Legacy Surveys $m_z<22.5$ galaxy sample. We construct a synthetic integrated-galaxy-light (IGL) map by depositing each source's $z$-band flux onto the HEALPix grid, producing a diffuse map with a well-defined input tomographic signal. Since the preceding source-mode tests already establish accurate redshift recovery for these resolved objects, the corresponding \tomographer\ output serves as the reference truth for the map-based measurement. We then degrade the map in two steps to emulate real intensity-map data: smoothing with a $5\arcmin$ Gaussian beam and adding a bright, spatially structured foreground constructed from Galactic polycyclic aromatic hydrocarbon (PAH) emission traced by the WISE 12~$\mu$m band \citep{2014ApJ...781....5M}, scaled to dominate the total map signal.

Figure~\ref{fig:Syn_IM} shows the \tomographer\ outputs for the high-resolution anchor map and the two progressively more realistic synthetic IGL maps. For the beam-smeared map, the effects are handled automatically by \tomographer\ through adaptive scale selection and the corresponding matter-clustering correction (Section~\ref{subsec:scales}). For the final, foreground-contaminated map, which retains the same beam smoothing, we additionally apply a $16^\circ$ high-pass filter and built-in template regression using the corrected-Schlegel--Finkbeiner--Davis (CSFD) Galactic reddening map \citep{chiang23}. As the imposed foreground is WISE 12~$\mu$m PAH emission, the mismatch with the CSFD template reflects the realistic situation in which the true foreground is not known exactly. The high-pass filter suppresses large-scale systematics, while template regression reduces smaller-scale Galactic foreground contamination. For both synthetic maps, the recovered $b\,({\rm d}I/{\rm d}z)$ remains consistent with the reference within the uncertainties, validating the intensity-mapping mode of \tomographer\ under realistic angular-resolution and foreground conditions.

In general, Galactic foregrounds should be statistically uncorrelated with the extragalactic LSS and therefore not bias clustering-redshift measurements. Template cleaning is thus not always necessary, and high-pass filtering alone may suffice. However, bright foregrounds can still introduce systematics through chance spatial correlations. We therefore recommend convergence tests across sky regions and foreground environments, verifying that the recovered $b\,({\rm d}I/{\rm d}z)$ remains stable where contamination is sufficiently controlled.

\section{Example Applications}\label{sec:examples}

Having established the \tomographer\ methodology, implementation, and validation, we now return to the ten applications showcased in Figures~\ref{fig:source_examples} and \ref{fig:IM_examples}: five source catalogs with diverse selections and five intensity maps spanning the electromagnetic spectrum from radio to X-rays. \tomographer\ applies to any dataset tracing LSS, regardless of wavelength, source type, or whether the signal is resolved or diffuse. Each result is obtained from positional information alone, with a typical runtime of only a few minutes per sample. We then apply \tomographer\ to the full set of four Dark Energy Spectroscopic Instrument (DESI) photometric target classes, demonstrating how the redshift response of a spectroscopic galaxy survey could have been characterized before a single spectrum is obtained, opening the door to iterative refinement of target selection.

\subsection{Source Catalogs} \label{subsec:ex_sources}

Figure~\ref{fig:source_examples} presents five source-catalog applications selected by magnitude, color, photometric redshift, morphology, and variability, described below in order from the top to bottom panels.

\emph{Magnitude-limited samples.} For SDSS photometric galaxies \citep{2000AJ....120.1579Y} selected at $r<20$, 21, and 22, the recovered distributions progressively extend to higher redshift and increase in normalization for fainter samples, as expected. This illustrates the baseline application of clustering redshifts: empirically recovering the redshift distribution of a photometric catalog directly from source positions, without requiring photometric redshifts or follow-up spectroscopy.

\emph{Color-selected samples.} Broadband color or color--magnitude selection is widely used to isolate galaxy populations over targeted redshift ranges for cosmology and galaxy-evolution studies. DESI photometric ELG targets \citep{2023AJ....165..126R} provide a prominent example from a Stage IV dark-energy experiment. Using only the catalog R.A. and decl. coordinates, the recovered clustering-redshift distribution shows the expected primary population at $z\sim1$, together with a distinct low-redshift interloper population that is difficult to quantify from color selection alone. It also reveals a high-redshift tail beyond $z=1.6$, where the [O\,II] emission-line doublet shifts outside the DESI spectral coverage and can no longer be spectroscopically confirmed, yet remains evident through clustering redshifts. We apply this analysis to the full set of DESI target classes in Section~\ref{subsec:ex_desi}.

\emph{Photometric-redshift bins.} Historically, validating photometric-redshift distributions was among the primary applications envisioned for clustering-redshift techniques, particularly for weak-lensing cosmology but also photometric galaxy surveys more broadly. Here, Legacy Surveys galaxies are divided into 10 photo-$z$ bins of width $\Delta z_{\rm phot}=0.1$ based on the estimates of \citet{2022MNRAS.512.3662D}. The recovered distributions show that these photo-$z$'s provide reliable relative ordering in redshift while exhibiting significant scatter, asymmetric tails, and catastrophic outliers. They also reveal variations in performance across bins, including substantial broadening at higher redshifts as the precision degrades. Such fine-grained diagnostics require no SED assumptions and take only a few minutes per bin, making \tomographer\ well suited to optimizing, validating, and recalibrating photometric redshifts for current and future imaging surveys.

\emph{Morphology-selected low-redshift galaxies.} The Extending the Satellites Around Galactic Analogs Survey \citep[xSAGA;][]{2022ApJ...927..121W} identifies low-redshift galaxy candidates using a convolutional neural network (CNN) trained on multiband imaging, with spectroscopic labels from \citet{2021ApJ...907...85M}. Selecting this local population is essential for identifying faint satellite candidates around nearby, spectroscopically confirmed hosts. The classifier is driven primarily by galaxy morphology while also incorporating color information. We progressively select subsamples by the CNN-assigned probability of being at $z<0.03$ (colored symbols; sample sizes are given in the legend). The recovered clustering-redshift distributions shift progressively lower with increasing CNN probability, independently confirming that the machine learning classifier tracks the intended redshift selection. The trend measured by \tomographer\ is also consistent with the independent validation of \citet{2023ApJ...954..149D}.

\emph{Variability-selected AGN.} For Gaia variability-selected AGN candidates \citep{2023A&A...674A..41G}, \tomographer\ recovers a broad redshift distribution tracing the cosmic history of black hole growth, peaking near $z\sim1$ and extending to $z\gtrsim3$. Further splits by variability amplitude and structure-function shape, a proxy for variability timescale, produce only minor changes in the recovered distributions. This suggests that these time-domain properties of AGN are less strongly correlated with redshift than properties such as colors, and largely reflect intrinsic stochasticity. More broadly, this example highlights the applicability of \tomographer\ to time-domain source catalogs, including AGN and potentially supernovae and other transients, which are becoming increasingly important as the Vera C. Rubin Observatory Legacy Survey of Space and Time \citep{2019ApJ...873..111I} begins its decade-long survey.

\subsection{Intensity Maps} \label{subsec:ex_IM}

Figure~\ref{fig:IM_examples} presents five intensity-map applications for the EBL across the electromagnetic spectrum, with \tomographer\ recovering the redshift distribution of diffuse emission without individual source detection. We describe the panels below from top to bottom.

\emph{Optical IGL.} We first revisit the synthetic optical IGL map constructed by summing resolved Legacy Surveys sources in Section~\ref{subsec:val_IM}. The recovered $b\,({\rm d}I/{\rm d}z)$ accurately reproduces the input redshift distribution despite beam smearing and strong Galactic foregrounds, demonstrating a seamless transition from source catalogs to diffuse intensity maps.

\emph{Far-infrared CIB.} The Planck 857\,GHz map \citep{2016A&A...594A...8P} recovers the broad redshift distribution of the CIB, peaking at $z\sim1$--2 and closely tracing the cosmic star-formation history \citep{2014ARA&A..52..415M}. It forms one of the 11 frequency bands used in the CIB tomography analysis of \citet{chiang25}, which provides one of the most precise and direct measurements to date of the cosmic star-formation history and the evolution of the cosmic dust density, $\Omega_{\rm dust}$, over $0<z<4$.

\emph{Microwave Compton $y$.} The Atacama Cosmology Telescope (ACT) Planck internal linear combination (ILC) Compton-$y$ map \citep{2024PhRvD.109f3530C} (red) and Planck needlet internal linear combination (NILC) Compton-$y$ map \citep{2016A&A...594A..22P} (blue) both recover the low-redshift thermal SZ signal together with a distinct high-redshift component consistent with residual CIB leakage. In contrast, \citet{2020ApJ...902...56C} separated the thermal SZ and CIB through multiband tomography with redshift-dependent component separation; the resulting best-fit SZ background is shown as the dashed–dotted curve, tracing the buildup of hot halo gas in the Universe. Because Compton $y$ measures thermal gas pressure, and hence energy, a robust CIB-free SZ tomography provides a census of the cosmic thermal history driven by structure formation. This example highlights clustering redshifts as a direct validation tool for current and future Compton-$y$ reconstructions. Importantly, such residual CIB contamination can also introduce systematic biases into $y$-autocorrelation measurements, as discussed by \citet{2025MNRAS.540.1055E}.

\emph{Diffuse X-ray background.} The ROSAT $0.1$--$2.4$\,keV diffuse map \citep{1997ApJ...485..125S} reveals two distinct components in redshift: AGN emission around the epoch of peak black hole growth ($z\sim1$--2) and a low-redshift upturn likely arising from hot halo gas in galaxy groups and clusters, which is also traced by the SZ effect in the previous $y$-map application. This illustrates how clustering redshifts can separate physically distinct components that are blended in the projected diffuse background.

\emph{Diffuse radio background.} Applied to the Murchison Widefield Array (MWA) 139--170\,MHz map from the GaLactic and Extragalactic All-sky MWA (GLEAM) survey \citep{2017MNRAS.464.1146H}, \tomographer\ recovers a broad radio AGN distribution peaking at $z\sim1$--2, similar to both the main X-ray component in the diffuse ROSAT background and the resolved Gaia AGN candidates shown earlier. The radio signal is dominated by synchrotron emission, and its common redshift evolution with the X-ray and optical AGN populations provides a complementary view of supermassive black hole growth across cosmic time. More broadly, redshift tomography offers a direct test of anomalous diffuse backgrounds such as the ARCADE~2 radio excess at lower frequencies \citep{2011ApJ...734....5F}: an extragalactic EBL component associated with LSS should correlate with other tracers and therefore be detected in clustering-redshift measurements, whereas an uncorrelated excess would instead point toward foregrounds or other systematics. A similar clustering-redshift test of the UV background \citep{2019ApJ...877..150C} disfavored an extragalactic origin for the mysterious UV monopole excess reported by the Galaxy Evolution Explorer \citep[GALEX;][]{2015ApJ...798...14H} and later New Horizons \citep{2025AJ....169..103M}, favoring instead instrumental or local astrophysical origins \citep{2022PASP..134h4302K}.

\begin{figure}[h!]
    \begin{center}
         \includegraphics[width=0.465\textwidth]{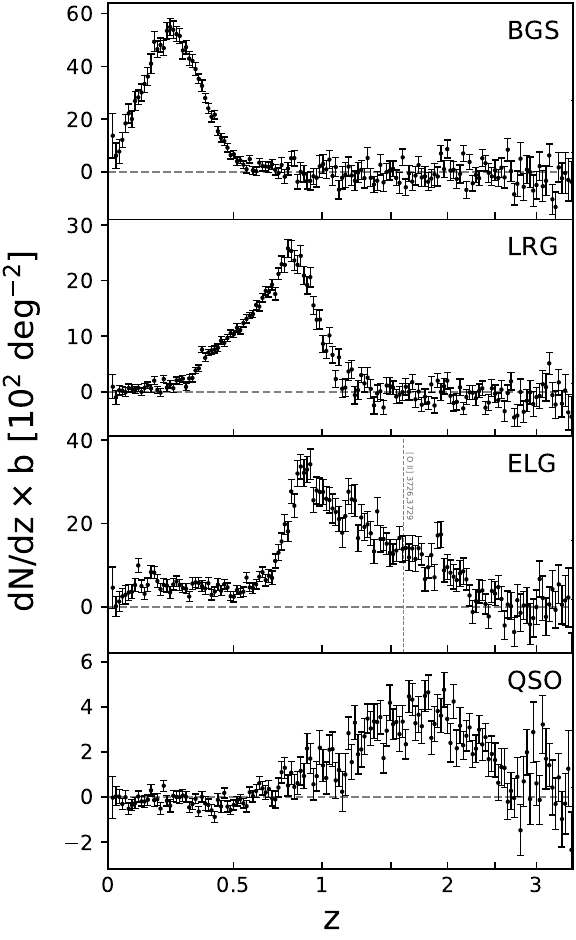}
    \end{center}
    \caption{Bias-weighted redshift distributions of the four DESI imaging target classes (BGS, LRG, ELG, and QSO), measured by \tomographer\ using the imaging catalogs alone. The recovered distributions resolve the sharp LRG cutoff at $z\sim1.1$, the broad low-redshift interloper plateau and high-redshift tail of the ELG sample, and the broad QSO distribution extending to $z\sim3$.}
    \label{fig:DESI_targets}
\end{figure}

\subsection{DESI Targets} \label{subsec:ex_desi}

As a survey-scale demonstration, we use \tomographer\ to recover the bias-weighted redshift distributions of the four DESI target classes \citep{2016arXiv161100036D} selected from the Legacy Surveys photometry \citep{2019AJ....157..168D}, as shown in Figure~\ref{fig:DESI_targets}: Bright Galaxy Survey \citep[BGS;][]{2023AJ....165..253H}, LRG \citep{2023AJ....165...58Z}, ELG \citep{2023AJ....165..126R}, and QSO \citep{2023ApJ...944..107C}. Using the imaging catalogs alone, \tomographer\ recovers the defining features of each class: the low-redshift BGS peak; the LRG distribution with a sharp cutoff at $z\sim1$; the ELG low-redshift interloper population and extended high-redshift tail; and the broad QSO distribution extending to $z\sim3$. These measurements require no DESI spectroscopy and could  therefore have been obtained before the spectroscopic survey begins.

For LRG, in addition to the main peak, we recover a characteristic kink at $z\sim0.35$, consistent with the independent validation of \citet{2023AJ....165...58Z} and likely reflecting a transition in the color-selection boundaries as spectral features move between the optical bands.

The ELG sample is particularly illustrative. While the primary target population lies near $z\sim1$, the clustering-redshift measurement reveals a substantial tail extending beyond $z=1.6$. At this point, the [O\,II] doublet shifts beyond the DESI spectral coverage, so this high-redshift population cannot generally be confirmed by DESI spectroscopy unless Ly$\alpha$ is identified \citep{2026arXiv260512503C}. Clustering redshifts therefore provide a complementary view of the full redshift response of the photometric selection, including populations only partially accessible to the spectroscopic observations themselves.

More generally, rapid clustering-redshift measurements from imaging catalogs enable early validation and iterative optimization of target selections before large spectroscopic samples become available. This capability will be increasingly valuable for future wide-area spectroscopic surveys, including DESI-II and Spec-S5 \citep{2025arXiv250307923B}, where improved target selection directly translates into greater survey efficiency and scientific return.

\section{Discussion}
\label{sec:discussion}

\subsection{Interpreting the Bias-weighted Distribution}
\label{subsec:bias_weighted}

The direct observable of clustering-redshift measurements is $b\,({\rm d}N/{\rm d}z)$, not ${\rm d}N/{\rm d}z$. \tomographer\ deliberately reports this quantity without attempting to remove $b$. When the total $N$ of the sample is known, the overall bias normalization can in principle be fixed by requiring ${\rm d}N/{\rm d}z$ to integrate to $N$; the analogous argument applies to a foreground-free intensity field with known mean intensity. What remains unconstrained is therefore primarily the redshift evolution of $b$. This evolution cannot be determined from either the cross-correlation data alone or a full-sample projected autocorrelation, and correcting for it would introduce model assumptions into an otherwise data-driven estimate.

For many applications, however, the bias weighting is benign. Narrow redshift distributions, such as tomographic photo-$z$ bins or narrowband ELG selections, are only weakly affected because the evolution of $b$ can often be neglected over the relevant redshift range. Similarly, many localization questions, such as peak positions, edges, outliers, and the overall redshift extent, can be addressed without precise knowledge of $b$.

For many LSS applications, the bias-weighted distribution is itself the physically relevant quantity. When the test sample is used as a tracer of the matter density field, for example in cross-correlations with other galaxy populations, cosmic microwave background (CMB) lensing \citep{2020JCAP...05..047K}, or the integrated Sachs--Wolfe effect \citep[e.g.,][]{2026arXiv260719648K}, the predicted or measured signal depends directly on $b\,({\rm d}N/{\rm d}z)$ rather than ${\rm d}N/{\rm d}z$. In this sense, the bias-weighted distribution is the more fundamental redshift kernel for matter tracers in LSS cosmology.

For some applications, however, the shape of ${\rm d}N/{\rm d}z$ itself is of interest, particularly when characterizing the intrinsic properties, formation, and evolution of galaxies. In such cases, additional observational or theoretical information, including autocorrelations, can help characterize the bias evolution through direct modeling or iterative schemes such as that explored by \citet{2008ApJ...684...88N}. Since $b(z)$ may have complex evolution with more degrees of freedom than can be constrained by these data, practical treatments generally adopt a simple, smooth parameterization, e.g., a power law in $(1+z)$, informed by available measurements, simulations, or theoretical expectations \citep[e.g.,][]{2018MNRAS.477.1664G,2020JCAP...05..047K}. The resulting correction for bias evolution can be applied to the measured $b\,({\rm d}N/{\rm d}z)$, with its uncertainty propagated into ${\rm d}N/{\rm d}z$. For typical, smoothly evolving samples, the resulting uncertainty in summary statistics such as the mean redshift can be as small as the percent level \citep{2013arXiv1303.4722M}.

A more direct and empirical way to reduce sensitivity to bias evolution is to combine clustering redshifts with available per-source features, dividing the parent sample into subsamples that are more localized in redshift, as advocated by \citet{2013arXiv1303.4722M}. Such subsamples can be defined using color--magnitude cuts or more general features, including methods such as self-organizing maps (SOMs) that exploit the available multiband photometry \citep{2015ApJ...813...53M}. The resulting narrower ${\rm d}N/{\rm d}z$ reduces the impact of bias evolution within each subsample, allowing their individually normalized distributions to be recombined with much less sensitivity to $b(z)$. The speed of \tomographer\ makes clustering-redshift measurements over large numbers of such finely divided samples practical.

\subsection{Systematics and Limitations}

\emph{Reference coverage.} \tomographer\ results are defined only over the redshift range and sky footprint of the reference sample, currently $0<z<4.2$ over $10{,}440\,{\rm deg}^2$ with SDSS. Unlike many clustering-redshift methods, \tomographer\ directly recovers the absolute normalization of $b\,({\rm d}N/{\rm d}z)$ within this span. Populations outside the reference redshift range are necessarily missed, while regions outside its footprint cannot contribute to the measurement. The sparse quasar reference at $z\gtrsim2.5$ (Figure~\ref{fig:ref_N_of_z_b_of_z}, bottom left) also increases statistical uncertainties of \tomographer\ toward the upper bound (e.g., Figure~\ref{fig:general_test_wide}). These limitations should be considered when normalizing the recovered ${\rm d}N/{\rm d}z$ into a probability distribution.

\emph{Test-sample systematics.} Spatial selection effects can, in general, modify both the normalization and shape of the recovered distribution. Users should therefore provide the most accurate available selection function via a random catalog or footprint mask. \tomographer\ further provides a direct empirical test through sky-region splits (Section~\ref{subsec:val_footprint}): a redshift-independent angular selection function rescales only the normalization, whereas more general spatial-redshift selection effects appear as changes in the recovered redshift-distribution shape. Foregrounds uncorrelated with the extragalactic LSS increase the uncertainties without biasing the recovered distribution (Section~\ref{subsec:val_IM}); users should nevertheless mitigate them using the built-in filtering and template-cleaning options in \tomographer\ and/or external preprocessing. The remaining concern is residual systematics shared by the test and reference samples but not captured by the random catalogs, e.g., residual dust-related selection effects \citep{2019ApJ...870..120C}; the latitude-split test in Section~\ref{subsec:val_footprint} indicates that such effects are negligible at the current statistical precision.

\emph{Magnification.} Gravitational lensing induces physical cross correlations between foreground and background populations: the test sample can magnify the reference sample, and vice versa, contaminating the wings of the recovered redshift distributions \citep{2013MNRAS.433.2857M,2018MNRAS.477.1664G,2020A&A...642A.200V}. The validation tests of Section~\ref{subsec:val_specz} place an empirical upper limit on this effect for the configurations considered here: no spurious signal is detected outside the true redshift extent at the current statistical precision. Applications targeting faint, highly magnified populations or precision measurements of the distribution wings should nevertheless account for magnification. Since the correction depends on the luminosity function of the input sample, it is inherently sample dependent and should therefore be applied by users in post-processing.

\emph{Scale dependence of bias.} Including nonlinear scales substantially boosts the SNR \citep{2013MNRAS.431.3307S}, at the cost of introducing scale-dependent bias that is absorbed into the effective $b_t$ and $b_r$. The fixed $\rp$ range with $\theta^{\gamma}$ weighting averages over scales identically for the auto- and cross correlations, so the leading effects cancel in Equation~(\ref{eq:master}). The agreement among reference tracers spanning a wide range of clustering bias (Figure~\ref{fig:b_ref_correction}), together with the independent analysis of \citet{2020JCAP...05..047K} (Figure~\ref{fig:unWISE}), indicates that any residual effects are empirically small. \tomographer\ also allows users to directly assess the impact of scale-dependent bias by repeating the analysis with different input beam FWHM values, which correspond to different clustering scales (Section~\ref{subsec:scales}),\footnote{For source catalogs, this can be achieved by first converting the catalog into a HEALPix source-density map and analyzing it in the intensity-map mode.} since the ratio between the recovered $b\,({\rm d}N/{\rm d}z)$ directly measures the relative scale dependence of the effective bias.

\subsection{Comparison to Existing Tools}

Public clustering-redshift codes, including \textsc{The-wiZZ} \citep{2017MNRAS.467.3576M}, \texttt{BallTreeXcorrZ} \citep{2020JCAP...05..047K}, and \texttt{yet\_another\_wizz} \citep{2020A&A...642A.200V}, implement the same underlying statistical approach and have enabled rigorous calibration programs in DES, KiDS, and elsewhere. \tomographer\ differs primarily in scope and implementation: (i) the package includes a curated, bias-characterized reference sample, removing a major practical burden; (ii) its activation-map architecture eliminates user-side pair counting, reducing runtimes from CPU hours to minutes or seconds; (iii) it operates on intensity maps as well as source catalogs; and (iv) it provides an end-to-end analysis pipeline. The trade-off is reduced flexibility: users adopt the precomputed reference sample, pixelization, and scale choices. For precision-cosmology applications requiring survey-specific references and covariance estimation, expert pipelines remain appropriate; \tomographer\ instead fills the complementary niche of fast, reliable, low-barrier redshift estimation for arbitrary source catalogs and intensity maps.

\subsection{Outlook}

The statistical power of clustering redshifts grows with the reference sample, giving the technique a particularly promising outlook as spectroscopic surveys accumulate. In the conventional photometric-redshift approach, improving performance generally requires deeper photometry, more bands, and broader wavelength coverage. These per-source requirements become increasingly demanding for fainter populations, with diminishing returns near practical limits set by sky noise, blending, and photometric calibration. Clustering redshifts scale differently and cumulatively: a growing spectroscopic reference can be repeatedly applied to newly detected test-sample populations, including sources far fainter than the reference galaxies and diffuse emission. Looking further ahead, there may be value in designing future spectroscopic surveys with clustering-redshift applications in mind, building toward a dense reference sample covering most of the extragalactic sky and an ever larger fraction of the observable Universe. Such a resource could provide lasting infrastructure for clustering-redshift measurements of essentially any extragalactic population.

Toward this broader vision, the near-term upgrade to \tomographer\ will come from DESI \citep{2016arXiv161100036D}. Although Data Release 1 is already publicly available \citep{2026AJ....171..285D}, its somewhat fragmented footprint is not yet ideal for constructing a contiguous reference, so we plan to begin incorporating DESI with Data Release 2 in a future \tomographer\ release. As the public DESI dataset grows, it will eventually increase the reference density by roughly an order of magnitude over much of the sky, with particularly large gains from the dense LRG and ELG samples at $0.5\lesssim z\lesssim1.6$ and improved quasar statistics up to $z\sim3.5$. The maximum reference redshift, however, will remain set by the SDSS quasars at $z\sim4.2$.

Because the activation-map architecture in \tomographer\ isolates the reference-data products from the user-facing computation, these upgrades require no changes to users' analysis workflows. Beyond DESI, the Subaru Prime Focus Spectrograph \citep{2014PASJ...66R...1T}, 4MOST \citep{2019Msngr.175....3D}, SPHEREx \citep{2026ApJ...999..139B}, Euclid \citep{2025A&A...697A...1E}, and Roman \citep{2015arXiv150303757S} will further extend the density, depth, and footprint of the reference sample, progressively expanding the reach of clustering-redshift measurements.

\section{Summary}\label{sec:conclusion}

We have presented \tomographer, an end-to-end clustering-based redshift estimation framework that brings the underlying methodology and implementation into a single package. By shifting the most computationally intensive spatial-correlation tasks into precomputed data products, \tomographer\ substantially lowers the technical barrier to clustering-redshift measurements, allowing most analyses to be completed in only a few minutes on a laptop. Its main results are as follows:

\begin{enumerate}
    \item \tomographer\ applies, without modification, to both source catalogs and diffuse intensity maps at any wavelength. It recovers the bias-weighted number-density redshift distribution, $b\,({\rm d}N/{\rm d}z)$, for source catalogs and the bias-weighted intensity redshift distribution, $b\,({\rm d}I/{\rm d}z)$, for intensity maps, the latter without requiring individual source detection.

    \item The core algorithm replaces traditional cross-correlation pair counting with simple dot products against precomputed activation maps on a fixed pixel grid for each redshift slice. As a result, the measurement cost is set by the processing configuration rather than the number of sources, effectively reducing the source-number scaling from $\mathcal{O}(N\log N)$ to $\mathcal{O}(1)$.
    
    \item The reference sample consists of about $3\times10^6$ SDSS spectroscopic galaxies and quasars over $10{,}000\,{\rm deg}^2$, spanning $0<z\lesssim4$. Its bias factor is measured in each redshift slice and divided out, making the recovered $b\,({\rm d}N/{\rm d}z)$ and $b\,({\rm d}I/{\rm d}z)$ independent of the chosen reference tracer.
    
    \item Validation tests using samples with known redshift distributions show robust recovery of $b\,({\rm d}N/{\rm d}z)$, resolving sharp bin edges at the redshift-binning resolution with no significant signal outside the true range. Bootstrap uncertainties yield unbiased, unit-variance Gaussian residuals. Further tests demonstrate robustness to footprint splits, spatially varying selection functions, beam smoothing, and bright foreground contamination.

    \item We demonstrate the broad applicability of \tomographer\ with five source-catalog applications spanning different selections and five intensity-map applications across the electromagnetic spectrum. We also compare the recovered unWISE galaxy distributions with \citet{2020JCAP...05..047K} and recover the redshift distributions of all four DESI target classes directly from imaging, before any DESI spectroscopy is used.
    
\end{enumerate}

\tomographer\ is publicly available on GitHub at \url{https://github.com/yuvoonng/tomographer}. As spectroscopic surveys map the LSS of the observable Universe with increasing density, depth, and sky coverage, clustering-redshift measurements will become correspondingly more powerful. The activation-map architecture allows \tomographer\ to incorporate these advances through future reference-sample updates while preserving the same user-facing workflow.

\begin{acknowledgments}
We thank Rongpu Zhou for providing the photometric DESI target catalogs, John Wu for providing the xSAGA satellite-galaxy catalogs, and Alex Krolewski for providing the unWISE clustering-redshift measurements. We also thank Yun-Ting Cheng for helpful feedback while beta-testing the \tomographer\ user interface. We acknowledge the use of data products provided by the CADE, a service of IRAP-UPS/CNRS, which supplied the HEALPix versions of several intensity maps used in this work. Y.-K.C., Y.V.N., and Y.-R.L. are supported by the National Science and Technology Council of Taiwan under grants NSTC 111-2112-M-001-090-MY3 and NSTC 114-2112-M-001-063-MY3, and by Academia Sinica through the Career Development Award AS-CDA-113-M01. This research made use of data from the Sloan Digital Sky Survey (\url{https://www.sdss.org}), the DESI Legacy Imaging Surveys, unWISE, Gaia, Planck, ACT, ROSAT, and the Murchison Widefield Array GLEAM survey. We are grateful to the teams, funding agencies, and data centers that made these public datasets and services available. We also acknowledge the SciServer platform \citep{2020A&C....3300412T} of the Institute for Data Intensive Engineering and Science (IDIES) at Johns Hopkins University for hosting a historical \tomographer\ web service.

\end{acknowledgments}

\software{\texttt{NumPy} \citep{2020NumPy-Array}, 
\texttt{NumExpr} \citep{2018zndo...1492916C},
\texttt{healpy} \citep{2005ApJ...622..759G, 2019JOSS....4.1298Z}, 
\texttt{SciPy} \citep{2020SciPy-NMeth},
\texttt{pandas} \citep{2024zndo..13819579T},
\texttt{scikit-learn} \citep{2011JMLR...12.2825P},
\texttt{matplotlib} \citep{2019zndo...2893252C},
\texttt{astropy} \citep{astropy:2013, astropy:2018, astropy:2022},
\texttt{tqdm} \citep{2020zndo...4293724D}, 
\texttt{Jupyter} \citep{2016ppap.book...87K}}

\appendix
\renewcommand{\thefigure}{\thesection\arabic{figure}}
\setcounter{figure}{0}

\section{Scale Weighting of Sky Pair Counts}
\label{app:weighting}

The scale-integrated clustering amplitudes used throughout this work are defined in Section~\ref{subsec:deprojection} as
\begin{equation}
\bar{w}=\int_{\theta_{\rm min}}^{\theta_{\rm max}}W(\theta)w(\theta)\,\mathrm{d}\theta,
\qquad\qquad
W(\theta)=\theta^\gamma,
\end{equation}
where the weighting function is applied to the angular correlation function $w$, measured in one-dimensional $\theta$ bins. In contrast, the activation-map implementation described in Section~\ref{subsec:algorithm} applies the weighting directly to individual sky pixel-reference pairs before the correlation estimator is formed. Defining the weighted pair sums
\begin{equation}
DD_{W'}=\sum_{(i,j)\in DD}W'(\theta_{ij}),
\qquad\qquad
DR_{W'}=\sum_{(i,j)\in DR}W'(\theta_{ij}),
\end{equation}
the corresponding Davis--Peebles estimator is
\begin{equation}
\bar{w}'+1=\frac{n_R}{n_D}\frac{DD_{W'}}{DR_{W'}}.
\end{equation}
Here, we derive the relation between the pair weight $W'(\theta)$ used in the implementation and the weighting function $W(\theta)$ in the formal definition of $\bar{w}$. In the continuum limit,
\begin{equation}
DD_{W'}\propto\int W'(\theta)\theta[1+w(\theta)]\,\mathrm{d}\theta,
\qquad\qquad
DR_{W'}\propto\int W'(\theta)\theta\,\mathrm{d}\theta,
\end{equation}
where the additional factor of $\theta$ arises from the annular area element, $\mathrm{d}\Omega\sim2\pi\theta\,\mathrm{d}\theta$. After the random normalization in the Davis--Peebles estimator introduced in Section~\ref{subsec:estimator}, the constant term cancels, leaving
\begin{equation}
\bar{w}'=\frac{\int W'(\theta)\theta w(\theta)\,\mathrm{d}\theta}
{\int W'(\theta)\theta\,\mathrm{d}\theta}.
\end{equation}
To reproduce the weighting adopted in Equation~(\ref{eq:weighting}), the pair weight must satisfy $W'(\theta)\theta=W(\theta)$, and hence $W'(\theta)=\theta^{\gamma-1}$. Substituting this expression gives
\begin{equation}
\bar{w}'
=\frac{\int\theta^\gamma w(\theta)\,\mathrm{d}\theta}
{\int\theta^\gamma\,\mathrm{d}\theta}
=\frac{\bar{w}}
{\displaystyle\int_{\theta_{\rm min}}^{\theta_{\rm max}}\theta^\gamma\,\mathrm{d}\theta},
\qquad\qquad
\bar{w}
=\left[\frac{\theta^{\gamma+1}}{\gamma+1}\right]_{\theta_{\rm min}}^{\theta_{\rm max}}\bar{w}'.
\end{equation}
The conversion therefore depends only on the adopted angular weighting power index and integration limits, and is independent of the functional form of $w(\theta)$.

\section{Reference Sample Sky Coverage}
\label{app:ref_mix}
\setcounter{figure}{0}

\begin{figure*}
    \gridline{\fig{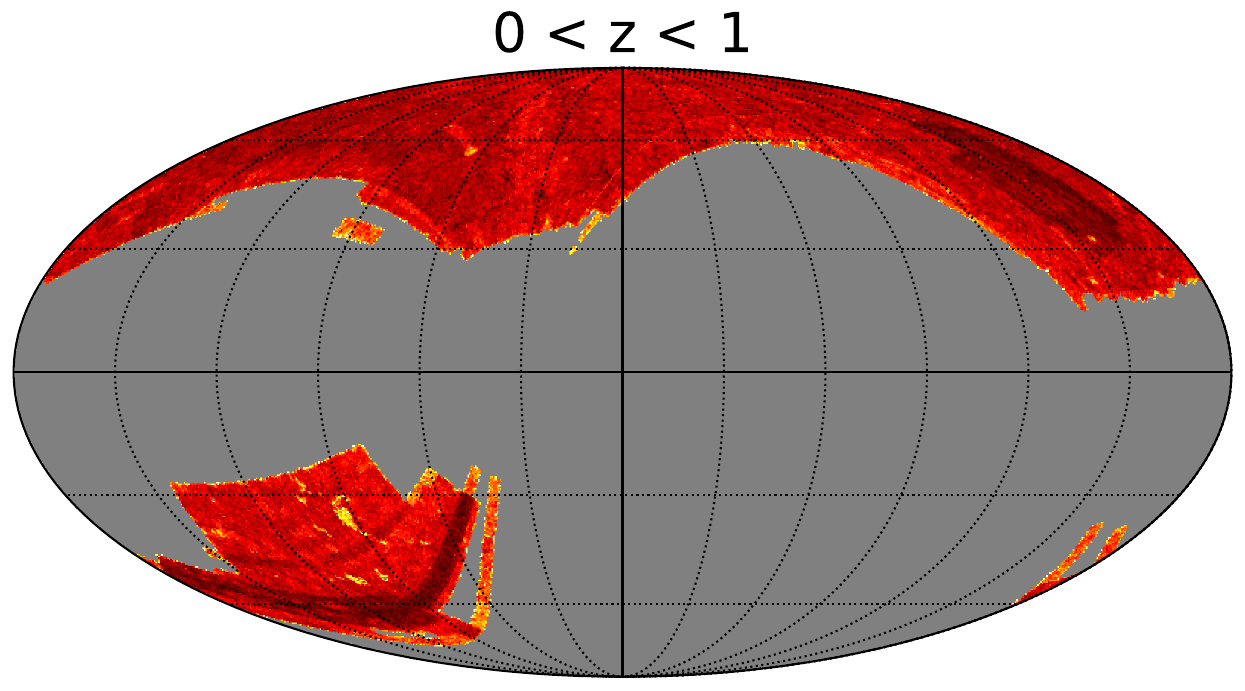}{0.33\textwidth}{}
              \fig{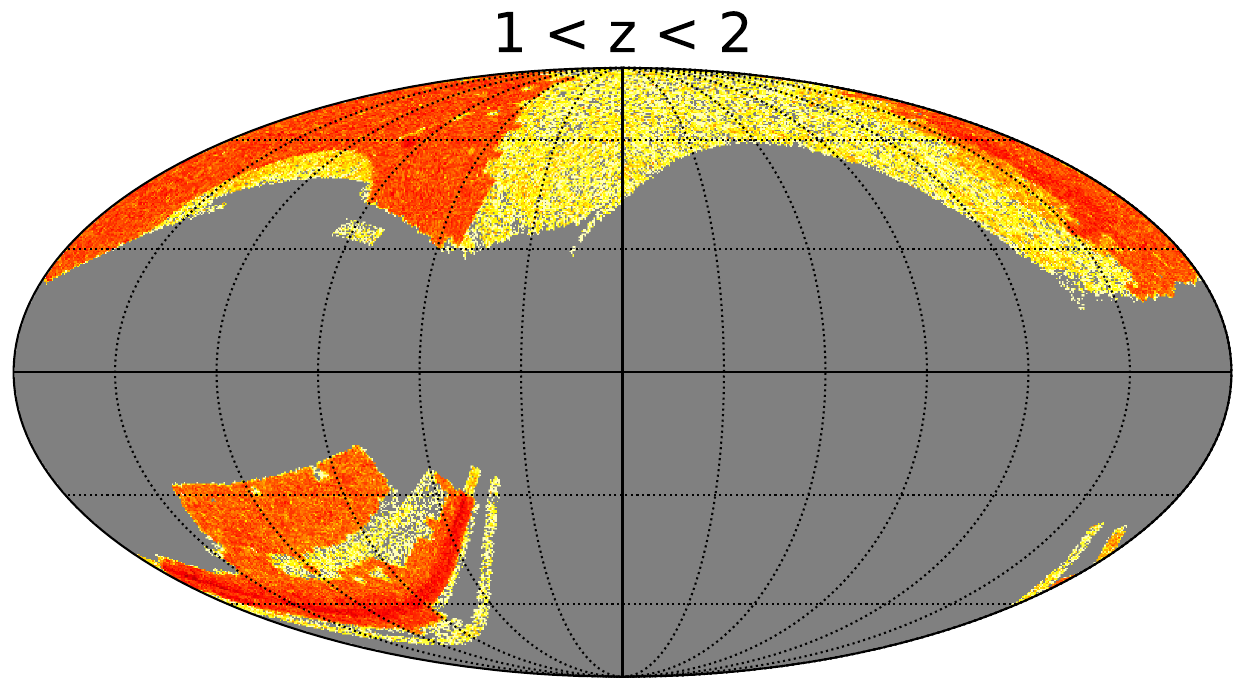}{0.33\textwidth}{}
              \fig{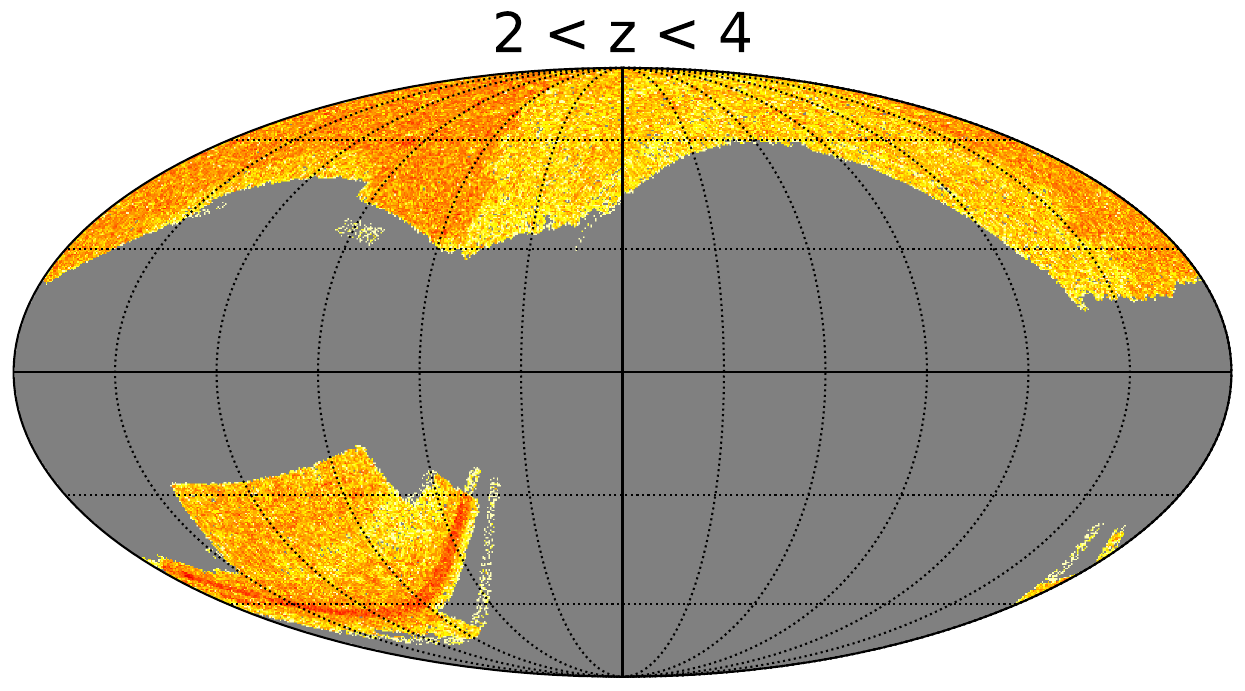}{0.33\textwidth}{}}
    \vspace{-1em}
    \gridline{\fig{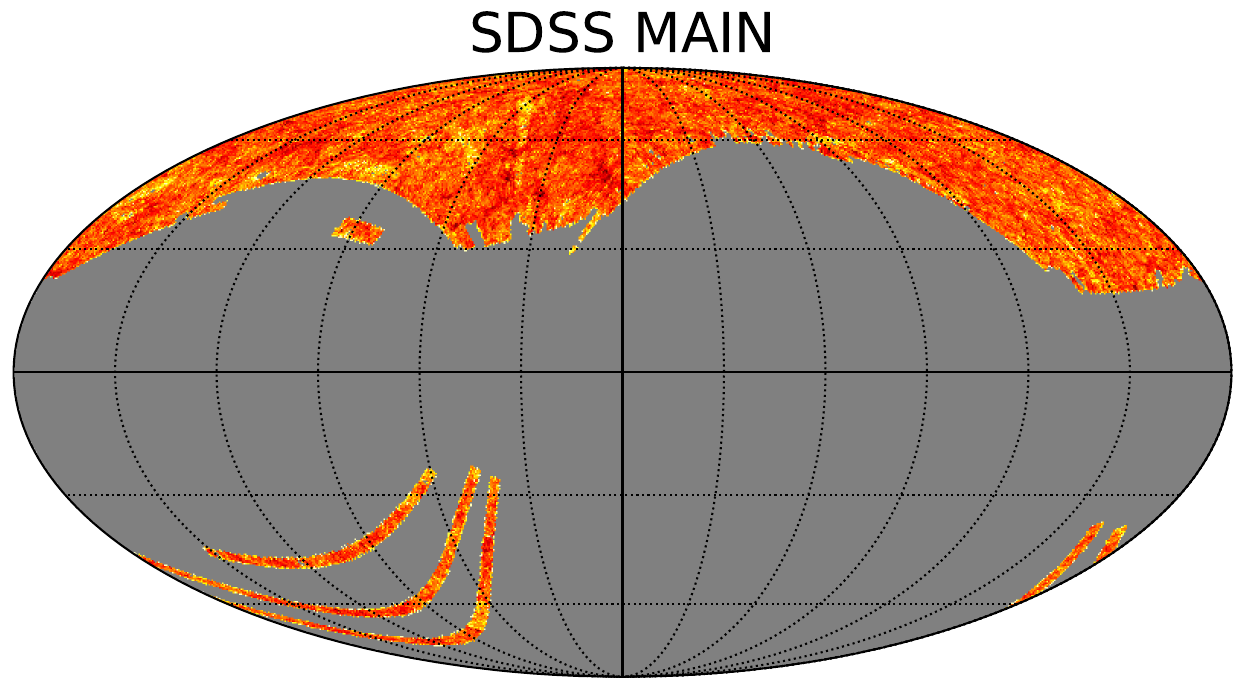}{0.33\textwidth}{}
              \fig{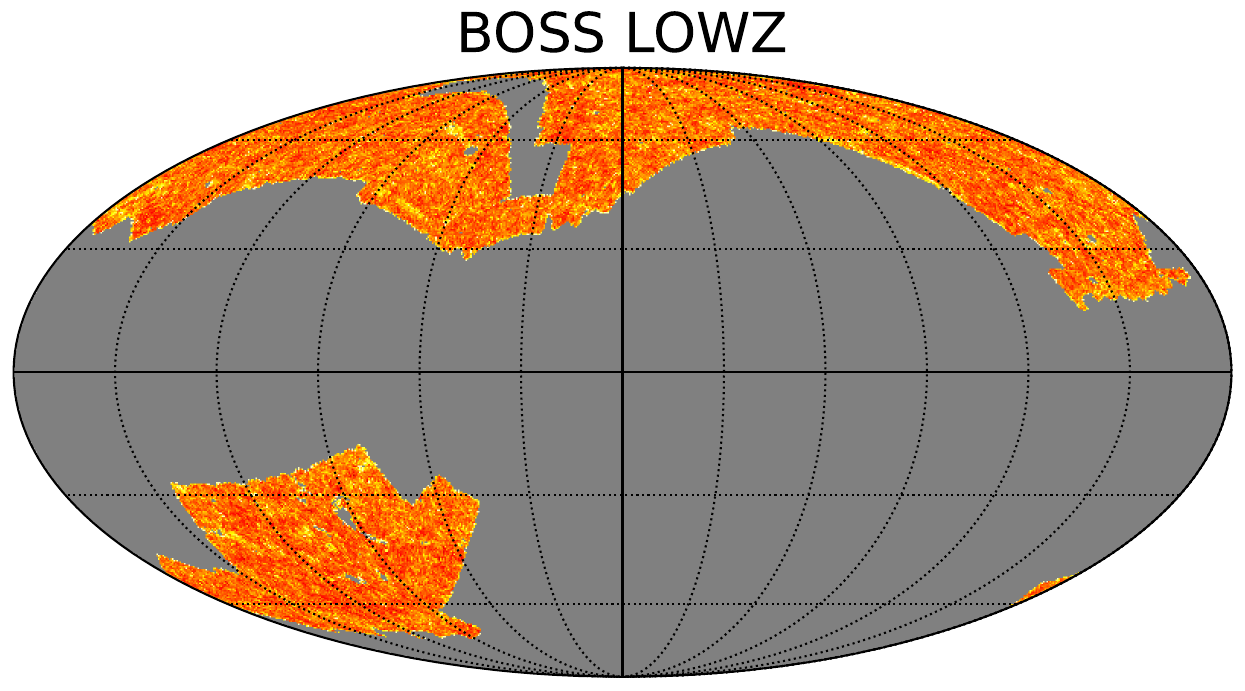}{0.33\textwidth}{}
              \fig{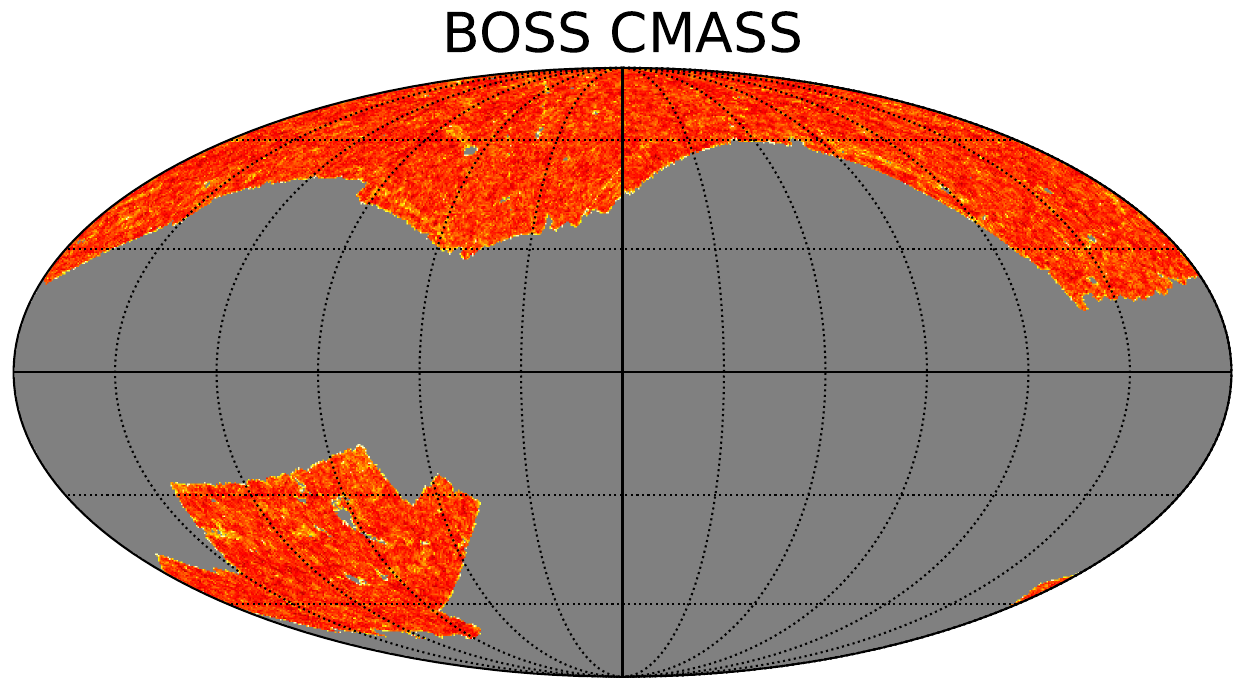}{0.33\textwidth}{}}
    \vspace{-2em}
    \gridline{\fig{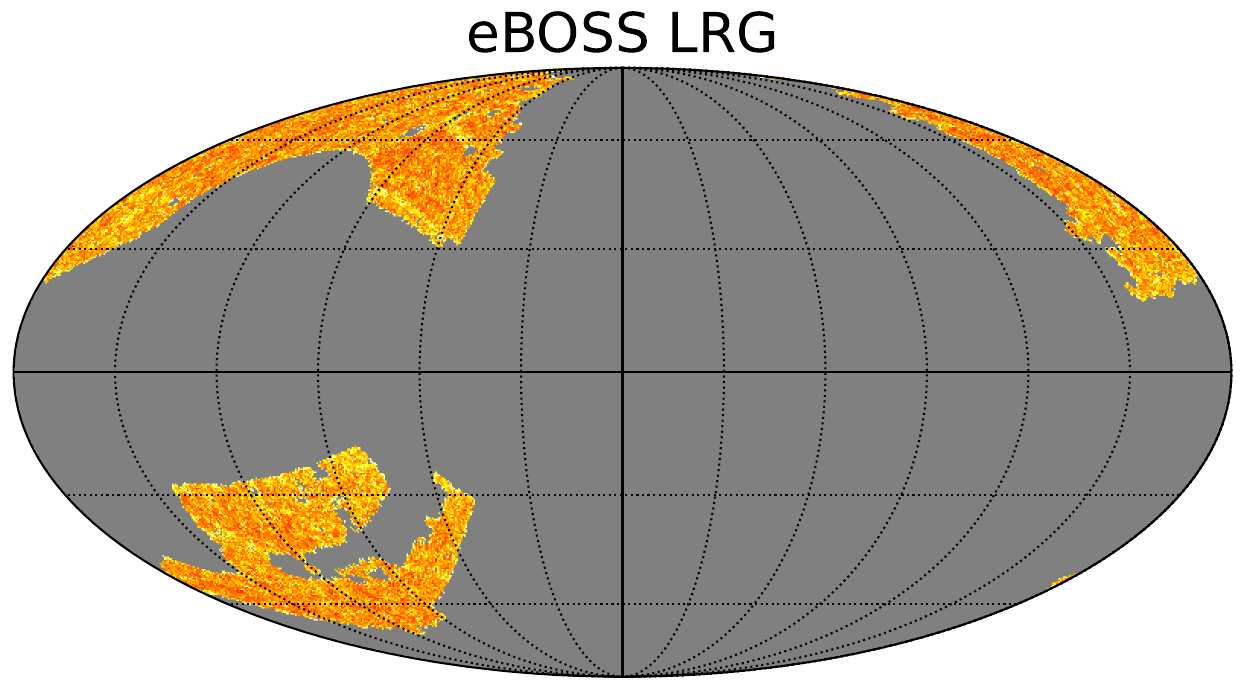}{0.33\textwidth}{}
              \fig{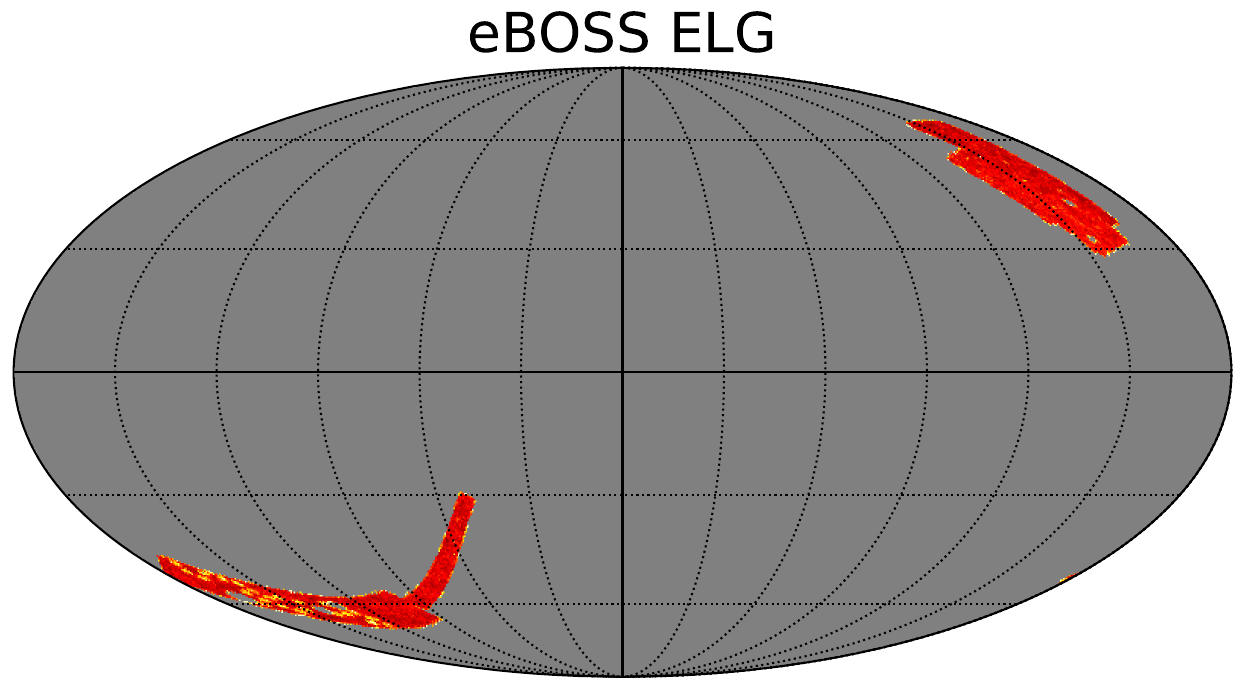}{0.33\textwidth}{}
              \fig{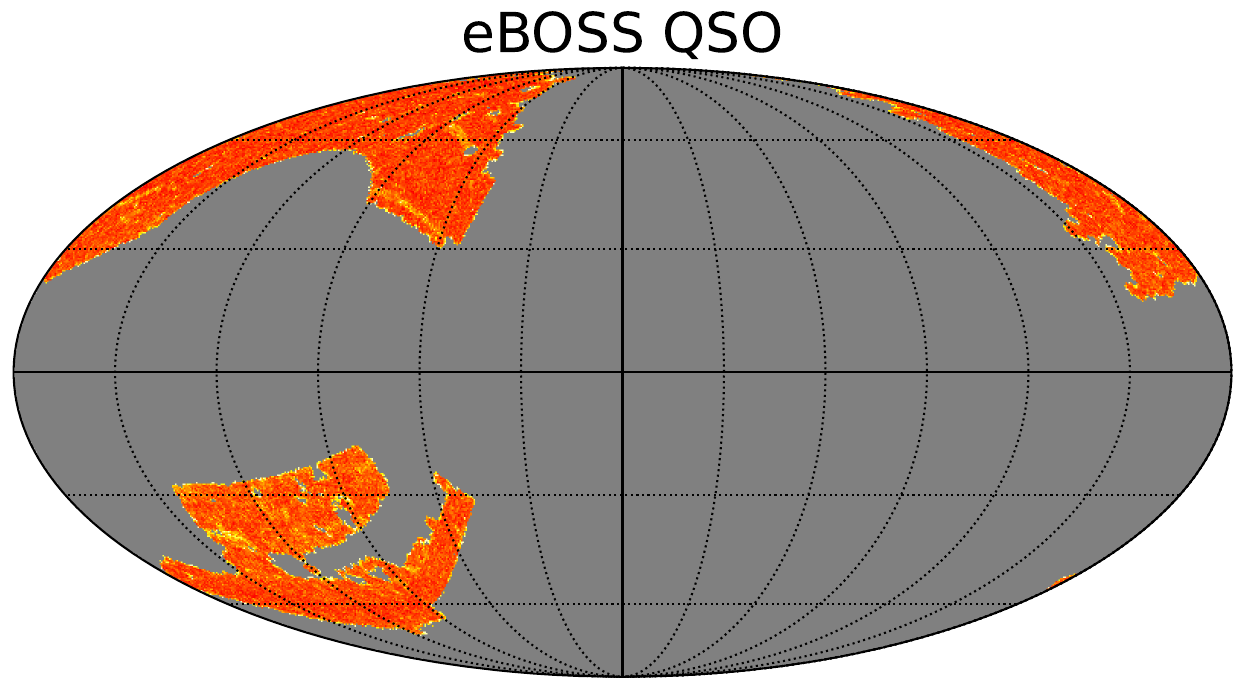}{0.33\textwidth}{}}
    \vspace{-2em}
    \gridline{\fig{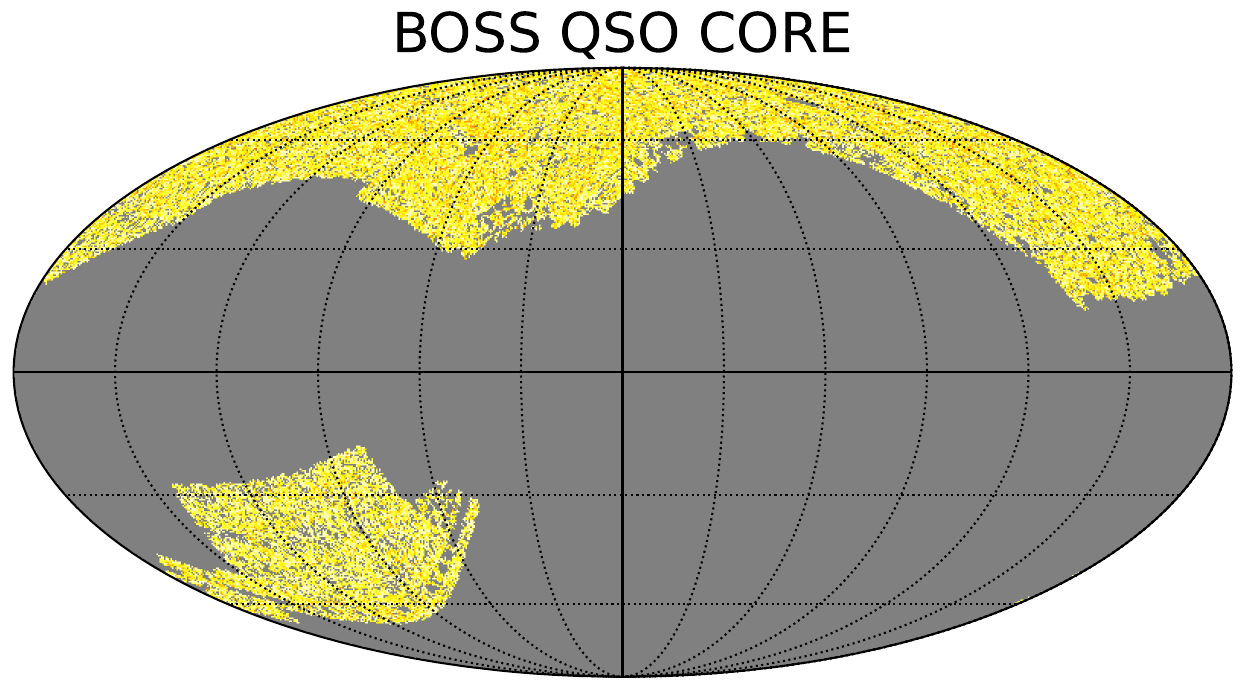}{0.33\textwidth}{}
              \fig{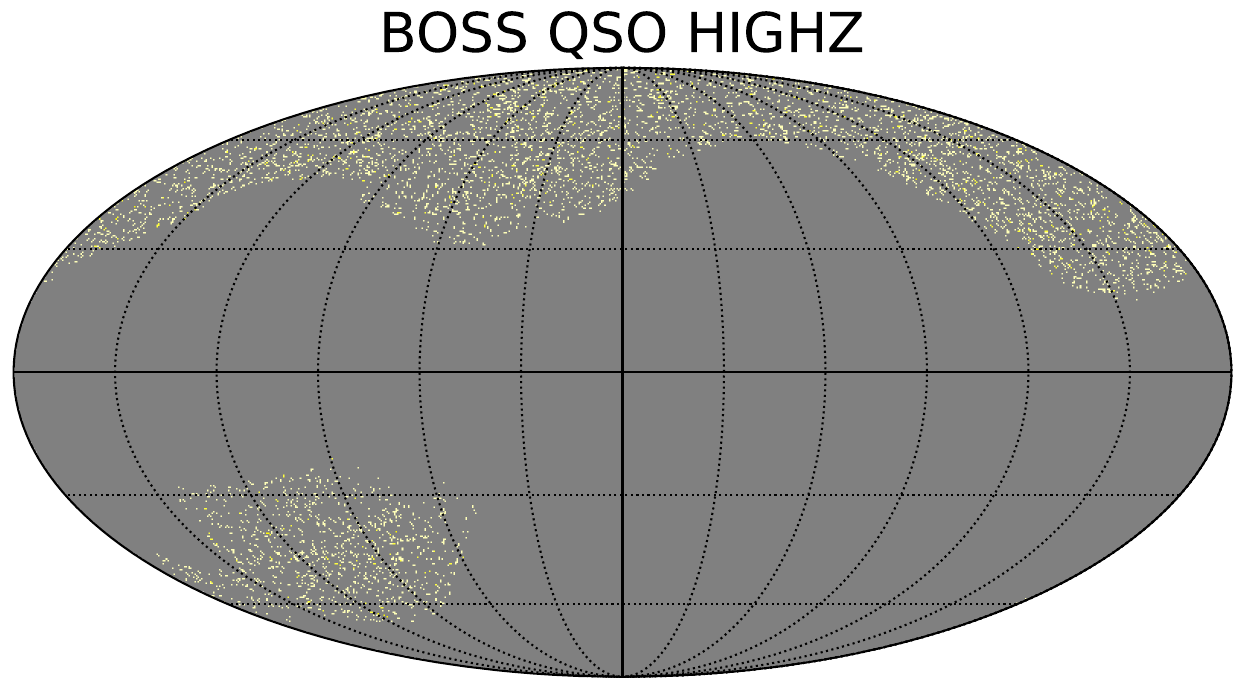}{0.33\textwidth}{}
              \fig{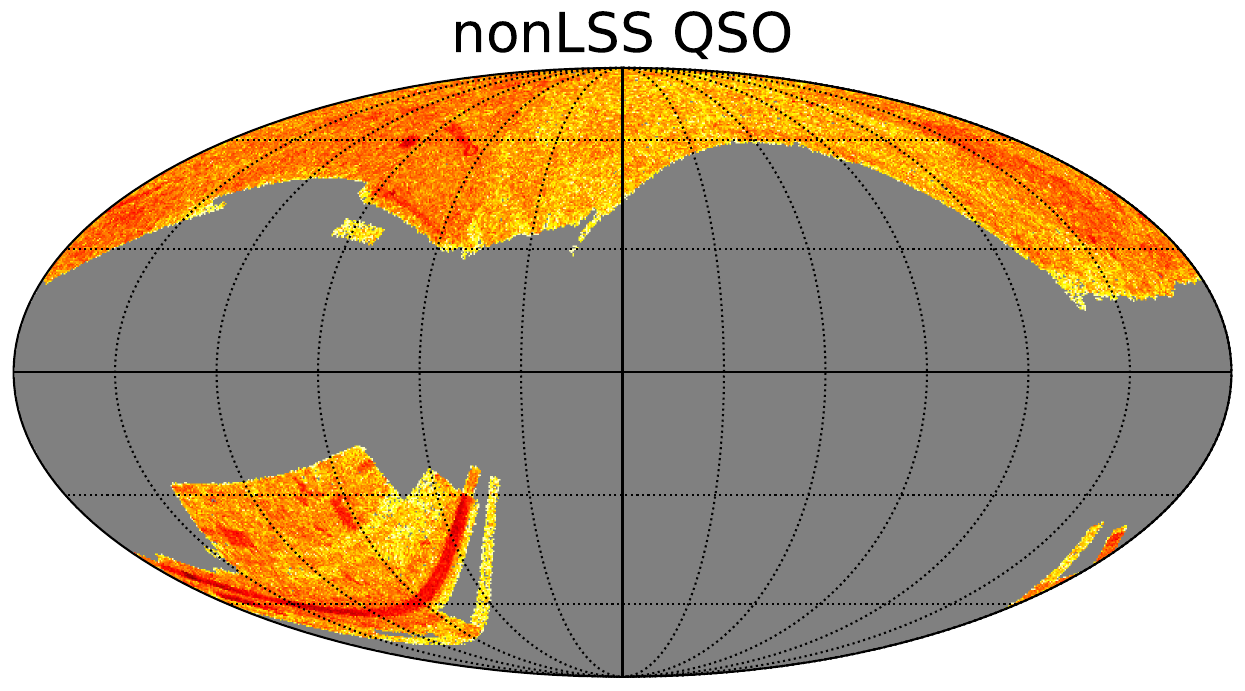}{0.33\textwidth}{}}
    \vspace{-2em}
    \gridline{\fig{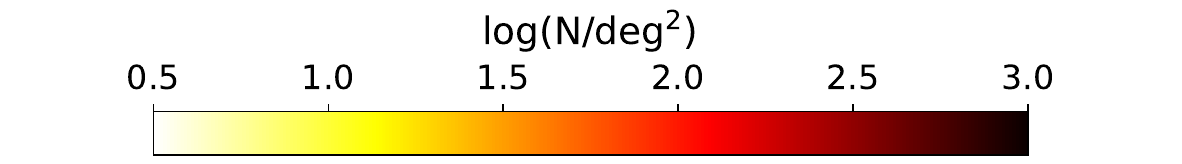}{0.55\textwidth}{}}

    \caption{Sky surface-density maps of the SDSS spectroscopic reference sample. The top row shows the combined sample in three broad redshift intervals ($0<z<1$, $1<z<2$, and $2<z<4$), dominated by galaxies at low redshift and quasars at higher redshifts. The remaining panels show the nine constituent subsamples listed in Table~\ref{tab:reference}. All panels share the same color scale in $\log(\#{\rm\,deg}^{-2})$.}
    \label{fig:SDSS_ref_by_sub_sample}
\end{figure*}

Figure~\ref{fig:SDSS_ref_by_sub_sample} summarizes the sky coverage of the nine SDSS spectroscopic reference subsamples listed in Table~\ref{tab:reference}. Together, they provide continuous redshift coverage over $0<z\lesssim4$, while illustrating the varying footprints, sky densities, and target populations that make up the combined reference sample.

\section{Built-in Masks, Foreground Templates, and Supplementary Data}
\label{app:built_in_masks}
\setcounter{figure}{0}

\begin{figure*}[!t]
    \begin{center}
         \includegraphics[width=1\textwidth]{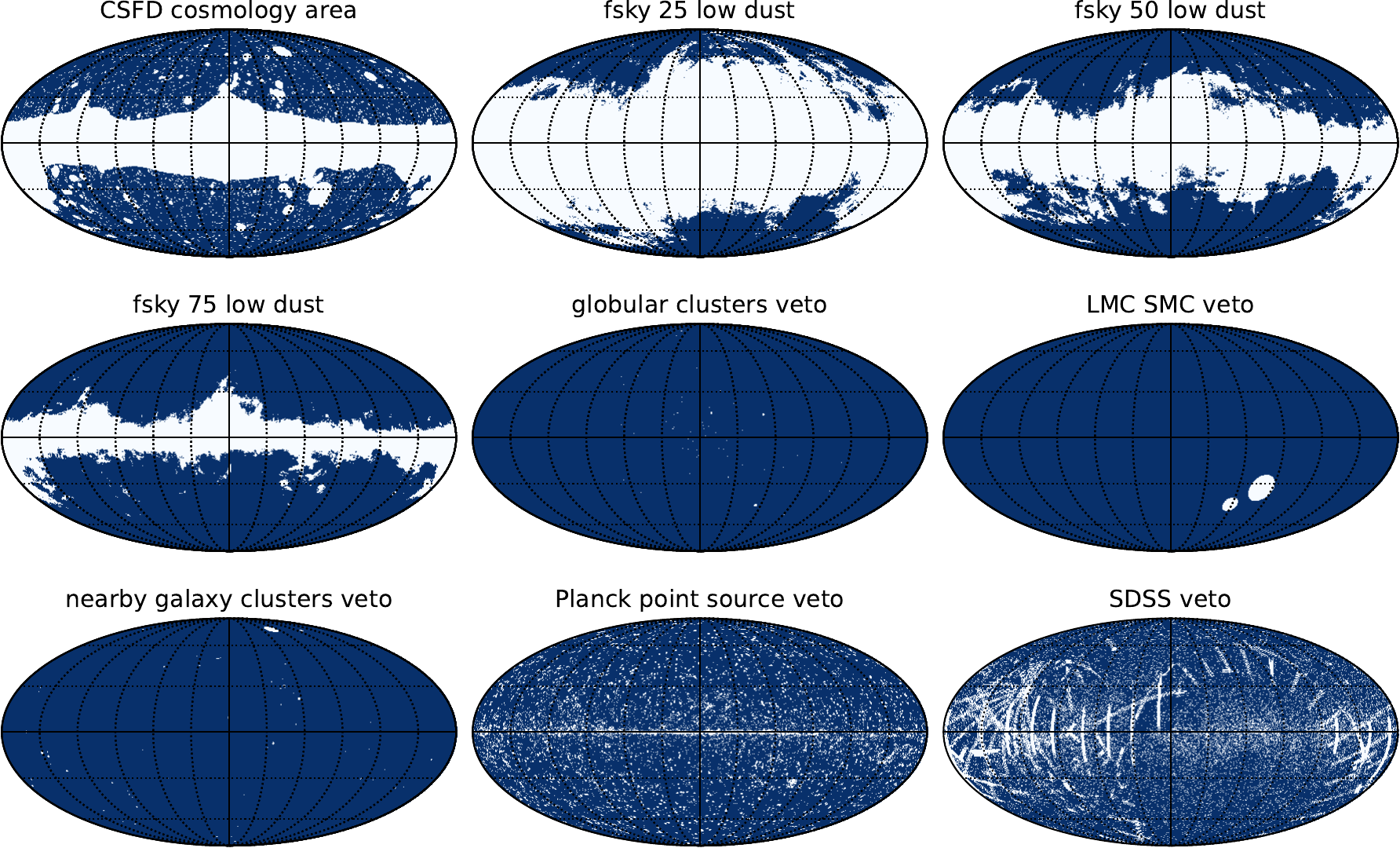}
    \end{center}
    \caption{Built-in masks distributed with \tomographer. These include the CSFD cosmology footprint, designed to retain clean high-latitude sky for extragalactic analyses; Galactic extinction masks; masks for the Magellanic Clouds, Galactic globular clusters, and nearby galaxy clusters; the \emph{Planck} compact-source catalog; and the SDSS imaging veto regions.}
    \label{fig:built_in_masks}
\end{figure*}

Figure~\ref{fig:built_in_masks} summarizes the built-in masking options provided by \tomographer. The available masks include the CSFD cosmology footprint \citep{chiang23}, designed to approximate the clean extragalactic sky for ultraviolet, optical, and near-infrared surveys; Galactic extinction masks retaining the cleanest 25\%, 50\%, and 75\% of the sky based on the same CSFD dust map; masks for the Magellanic Clouds constructed from Gaia stellar densities \citep{2016A&A...595A...1G}; Galactic globular clusters \citep{2013A&A...558A..53K}; nearby galaxy clusters from the Meta-Catalogue of X-ray detected Clusters \citep{2011A&A...534A.109P}; the \emph{Planck} compact-source catalog \citep{2016A&A...594A..26P}; and the SDSS imaging veto regions \citep{2005AJ....129.2562B}. These products provide convenient defaults for many analyses, although users may instead supply arbitrary $\mathrm{NSIDE}=2048$ HEALPix masks or weighting maps.

For Galactic template regression, \tomographer\ currently provides two built-in foreground templates. The CSFD reddening map \citep{chiang23} traces Galactic dust while minimizing CIB contamination. The HI4PI neutral hydrogen map \citep{2016A&A...594A.116H} traces diffuse Galactic H\,\textsc{i} and has also been shown to contain no detectable extragalactic LSS \citep{2019ApJ...870..120C}. These properties minimize the risk of removing the target cosmological signal during regression. Users wishing to employ alternative foreground tracers can perform the corresponding template regression externally before running \tomographer.

In Table~\ref{tab:zenodo}, we provide descriptions of the precomputed files for the full 160 tomographic bins available on Zenodo 
\citep{chiang_2026_20155554}, along with their corresponding roles in our algorithm. 

\begin{deluxetable*}{ccc}
\tablecaption{List of precomputed files uploaded to Zenodo \citep{chiang_2026_20155554}. $r_\mathrm{p,min} \in \{0.5,\, 1,\, 1.5,\, 2,\, 2.5\}$ denotes the minimum projected physical separations in Mpc. \label{tab:zenodo}}
\tablehead{\colhead{File} & \colhead{Description} & \colhead{Usage}}
\startdata
\addlinespace[5pt]
\makecell{\texttt{b\_ref\_sdss\_lss\_plus\_$r_
\mathrm{p,min}$-10Mpch}\\\texttt{\_gamma-0.8.fits}} & Reference bias $b_r$ from autocorrelations & On-the-fly $b_r$ with $\mathrm{SNR}<5$ \\
\addlinespace[5pt]
\midrule[0.1pt]
\addlinespace[5pt]
\texttt{sdss\_lss\_plus\_data\_z\_maps.pddf.parquet} & 
Pixelized sky maps for the reference sample & $\bar{w}_{rr}$\\
\addlinespace[5pt]
\midrule[0.1pt]
\addlinespace[5pt]
\makecell{\texttt{sdss\_lss\_plus\_data\_$r_
\mathrm{p,min}$-10Mpch}\\\texttt{\_gamma-0.8.pddf.parquet}} & Data activation maps (Sec.~\ref{subsec:algorithm}) & $\bar{w}_{rr},\,\bar{w}_{tr}$ \\
\addlinespace[5pt]
\midrule[0.1pt]
\addlinespace[5pt]
\makecell{\texttt{sdss\_lss\_plus\_rand\_$r_
\mathrm{p,min}$-10Mpch}\\\texttt{\_gamma-0.8.pddf.parquet}} & Random activation maps (Sec.~\ref{subsec:algorithm}) & $\bar{w}_{rr},\,\bar{w}_{tr}$ \\
\addlinespace[5pt]
\midrule[0.1pt]
\addlinespace[5pt]
\makecell{\texttt{wmbar\_prp\_$r_
\mathrm{p,min}$-10Mpch\_gamma-0.8}\\\texttt{\_15.0degtmax\_160log1pzbins}\\\texttt{\_multi\_beam\_filtering.fits}} & \makecell{Matter correlation function $\bar{w}_{m}$ \\ at multiple filtering scales} & Correction to $b_r$ and $\bar{w}_{tr}$\\
\addlinespace[5pt]
\enddata
\end{deluxetable*}

\bibliography{references}{}
\bibliographystyle{aasjournal}

\end{document}